\documentclass[fleqn,usenatbib]{mnras}

\usepackage{newtxtext,newtxmath}

\usepackage[T1]{fontenc}

\DeclareRobustCommand{\VAN}[3]{#2}
\let\VANthebibliography\thebibliography
\def\thebibliography{\DeclareRobustCommand{\VAN}[3]{##3}\VANthebibliography}

\usepackage{graphicx}	
\usepackage{amsmath}	

\usepackage{amssymb}	
\usepackage{tablefootnote}
\usepackage{multicol}

\newcommand{\kms}{\mbox{km s$^{-1}$}}

\newcommand{\muJb}{\mbox{$\mu$Jy~beam$^{-1}$}}
\newcommand{\Msolxyr}{\mbox{$M_{\sun}$ yr$^{-1}$}}

\title[AT2020vwl]{A radio-emitting outflow produced by the tidal disruption event AT2020vwl}

\author[A. J. Goodwin et. al.]{
A. J. Goodwin,$^{1}$\thanks{E-mail: ajgoodwin.astro@gmail.com} 
K. D. Alexander$^2$,
J. C. A. Miller-Jones$^1$,
M. F. Bietenholz$^3$, 
S. van Velzen$^4$,
G. E. Anderson$^1$,
\newauthor
E. Berger$^5$,
Y. Cendes$^5$,
R. Chornock$^6$,
D. L. Coppejans$^7$,
T. Eftekhari$^8$,
S. Gezari$^{9}$,
T. Laskar$^{10}$,
\newauthor
E. Ramirez-Ruiz$^{11}$,
and R. Saxton$^{12}$
\\
$^{1}$International Centre for Radio Astronomy Research -- Curtin University, GPO Box U1987, Perth, WA 6845, Australia \\
$^2$Department of Astronomy/Steward Observatory, University of Arizona, 933 North Cherry Avenue, Tucson, AZ 85721-0065, USA\\
$^3$Department of Physics and Astronomy, York University, Toronto,
M3J~1P3, Ontario, Canada \\
$4$ Leiden Observatory, Leiden University, Postbus 9513, 2300 RA, Leiden, The Netherlands \\
$^5$ Center for Astrophysics | Harvard \& Smithsonian, Cambridge, MA 02138, USA \\
$^6$Department of Astronomy, University of California, Berkeley, CA 94720-3411, USA\\
$^7$Department of Physics, University of Warwick, Coventry CV4 7AL, UK\\
$^{8}$Center for Interdisciplinary Exploration and Research in Astrophysics (CIERA) and Department of Physics and Astronomy, Northwestern University, Evanston, IL 60208, USA\\
$^{9}$Space Telescope Science Institute, 3700 San Martin Drive, Baltimore, MD  21211\\
$^{10}$Department of Astrophysics/IMAPP, Radboud University, PO Box 9010, 6500 GL, The Netherlands \\
$^{11}$Department of Astronomy and Astrophysics, University of California Santa Cruz, 1156 High Street, Santa Cruz, CA 95060, USA \\
$^{12}$Telespazio UK for ESA, European Space Astronomy Centre, Operations Department, 28691 Villanueva de la Cañada, Spain\\
}

\date{Accepted 2023-04-24. Received 2023-04-20; in original form 2023-03-24}

\pubyear{2023}

\begin{document}
\label{firstpage}
\pagerange{\pageref{firstpage}--\pageref{lastpage}}
\maketitle

\begin{abstract}

A tidal disruption event (TDE) occurs when a star is destroyed by a supermassive black hole. 
Broadband radio spectral observations of TDEs trace the emission from any outflows or jets that are ejected from the vicinity of the supermassive black hole. However, radio detections of TDEs are rare, with $<20$ published to date, and only 11 with multi-epoch broadband coverage. 
Here we present the radio detection of the TDE AT2020vwl and our subsequent radio monitoring campaign of the outflow that was produced, spanning 1.5 years post-optical flare. We tracked the outflow evolution as it expanded between $10^{16}$\,cm to $10^{17}$\,cm from the supermassive black hole, deducing it was non-relativistic and launched quasi-simultaneously with the initial optical detection through modelling the evolving synchrotron spectra of the event. 
We deduce that the outflow is likely to have been launched by material ejected from stream-stream collisions (more likely), the unbound debris stream, or an accretion-induced wind or jet from the supermassive black hole (less likely). AT2020vwl joins a growing number of TDEs with well-characterised prompt radio emission, with future timely radio observations of TDEs required to fully understand the mechanism that produces this type of radio emission in TDEs.

\end{abstract}




\section{Introduction}

Tidal disruption events (TDEs) occur when a star passes too close to a supermassive black hole (SMBH) at the center of a galaxy and is destroyed \cite[e.g.][]{Hills1975,Rees1988}. After the stellar disruption, approximately half of the stellar debris remains bound to the SMBH and is accreted, while the other half is ejected from the SMBH on hyperbolic orbits \citep[e.g.][]{Rees1988}. The bound material is thought to be the source of observed optical and X-ray emission \citep[e.g.][]{vanVelzen2020}.
Observations of TDEs \citep[see][for a review]{Gezari2021} enable direct measurements of accretion events onto SMBHs and the subsequent launching of jets and outflows that may be produced  \citep{DeColle2012,2018ApJ...865..128L,Alexander2020}. 

Radio emission from TDEs traces outflowing material that is ejected from the vicinity of the SMBH due to the stellar disruption \citep[see][for a review]{Alexander2020,2018ApJ...865..128L}, either due to collimated jets \citep[e.g.][]{Zauderer2011,Bloom2011,Burrows2011,Levan2011,Lei2016,vanVelzen2016,Pasham2022,Andreoni2022} or sub-relativistic, wide-angle outflowing material \citep[e.g.][]{Alexander2016,Cenko2016,Blagorodnova2019,Hung2019}. Recently there has been an increase in the number of radio-detected TDEs, with large radio observational campaigns targeting optical and X-ray selected events \citep[][]{Alexander2020,Alexander2021ATel}. To date, just a handful of relativistic jets have been observed from TDEs exhibiting non-thermal spectral properties \citep{Zauderer2011,Berger2012,Cenko2012,Zauderer2013,Brown2017,Eftekhari2018,Mattila2018,Cendes2021a,Wiersema2020,Pasham2022,Andreoni2022}, whilst radio detections of prompt non-relativistic outflows from TDEs exhibiting thermal spectral properties are becoming increasingly more common \citep[e.g.][and references therein]{Alexander2020,Stein2021,Cendes2021,Goodwin2022,Goodwin2023}. In some cases, the radio flare can be delayed by up to years post-optical flare \citep{Horesh2021,Horesh2021b,Cendes2022,Perlman2022}.

Whilst the energetic radio emission observed from relativistic TDEs is consistently explained by a relativistic jet launched from the SMBH \citep[e.g.][]{Bloom2011,Cenko2012,Pasham2022}, the mechanism that produces the lower energy radio emission observed in non-relativistic thermal events is still under debate. Possible scenarios include an outflow produced by disk winds or a sub-relativistic jet, launched due to early accretion onto the SMBH \citep[e.g.][]{Pasham2018,vanVelzen2016,Alexander2016}. 
This scenario requires prompt circularisation of the debris material in order to explain the observed early onset of many radio outflows. Alternatively, the non-relativistic outflows could be explained by material ejected by stream-stream collisions of the stellar debris in a "collision-induced outflow" \citep{Lu2020}. In this scenario, prompt circularisation is not required, as early outflows can be launched while the debris are still circularising. Finally, the observed radio outflows could be produced by the unbound portion of the tidal debris stream, which has typical velocities $\sim10^{4}$\,\kms\ in a concentrated cone close to the orbital plane \citep{Krolik2016,Yalinewich2019}. Discrimination among these scenarios has proved difficult due to the similarities in energy and velocity of the outflowing material in all three cases, motivating further detailed observations that track their evolution \citep{Mockler2021}.

Combining such observations with new insights from simulations of TDEs may also aid attempts to distinguish among these possible scenarios. Recent work suggests that debris circularisation, in particular the efficiency thereof, plays a crucial role in the multiwavelength emission that is produced \citep[e.g.][]{Ramirez2009,Hayasaki2013,Bonnerot2016,Hayasaki2016,Guillochon2014,Shiokawa2015,Skadowski2016,Liptai2019,Bonnerot2020,Mummery2020}. Incoming and outgoing debris stream collisions have been suggested to drive much of the accretion disk formation efficiency \citep[e.g.][]{Hayasaki2013}, as well as ejecting material in outflows \citep[e.g.][]{Lu2020}.  Recently, \citet{Steinberg2022} found that in 
3D radiation-hydrodynamical simulations, the lightcurve rise is initially,
up to the optical peak, powered by shocks due to inefficient circularisation of the debris, and then after the peak, the debris efficiently circularises. In this model, outflows are produced initially as the debris is circularising, but stronger outflows are powered once the debris is circularised and the emission is predominately accretion-powered post-optical peak. Both \citet{Steinberg2022} and \citet{Andalman2022} found in hydrodynamical simulations that rapid ($\lesssim70$\,d) circularisation of the stellar debris occurs. \citet{Metzger2022} modelled the long-term evolution of the resulting envelope post-rapid debris circularisation and found that a cooling-induced envelope contraction that delays significant accretion onto the SMBH could produce delayed X-ray and radio emission, as has been observed in some TDEs. These models provide important predictions about what is driving the multiwavelength emission in TDEs, notably the radio emission, which has shown
diverse behaviour across different TDEs, including the production
of late-time radio flares in some cases.

Here we present an extensive radio monitoring campaign that we conducted on the optically-discovered TDE AT2020vwl. In Section \ref{sec:observations} we present the observations and data reduction, in Section \ref{sec:results} we present the results, including the radio lightcurve and spectra for each epoch. In Section \ref{sec:modelling} we model the radio emission, assuming a synchrotron spectrum for the transient component, to infer physical properties of the outflow. In Section \ref{sec:discussion} we discuss the implications of the results and compare this TDE with others, 
and finally in Section \ref{sec:conclusion} we summarise the results and provide concluding remarks.

\section{Observations}\label{sec:observations}

AT2020vwl (also Gaia20etp, ZTF20achpcvt) was first discovered by the \textit{Gaia} Spacecraft on 2020 October 10 as an optical flare of $\sim1$\,mag above the quiescent galaxy flux, localised to the centre of the galaxy SDSS J153037.80+265856.8/LEDA 1794348 \citep[J2000 RA, Dec 15:30:37.800, +26:58:56.89,][]{Gaia_ATel}. The event was also observed by the Zwicky-Transient Facility (ZTF) on 2020 October 8, but was not reported until 2020 December 20 \citep{ZTF_ATel,Yao2023}. A follow-up observation with the SED Machine integral field unit Spectrograph at the Palomar 60-inch on 2020 December 21 showed a spectrum with a steep blue optical continuum and strong, broad H, HeII, and Balmer lines \citep{ZTF_ATel} at a redshift of 0.0325, which corresponds to a luminosity distance, $D_L$ = 147\,Mpc, and an angular-size
distance $D_A$ of 138\,Mpc.
\citet{ZTF_ATel} classified the event as a H+He tidal disruption event based on the spectral properties and the bright UV flux measured by the Neil Gehrels Swift telescope. AT2020vwl's optical properties are typical of the optical TDE population \citep{Yao2023}. Like most optical TDEs, AT2020vwl did not have X-ray emission detectable in the early \textit{Swift} observation \citep{ZTF_ATel}. 

\subsection{VLA}

We obtained seven epochs of radio observations of AT2020vwl with the NRAO's Karl G. Jansky Very Large Array (VLA) from 2021 February 23 to 2022 May 08 across 1--18\,GHz (L- to Ku-band), via our VLA large program
to follow-up TDEs within $z<0.1$\footnote{\url{https://www.as.arizona.edu/radiotdes}} (program ID 20B-377, PI: Alexander). On 2021 February 23 we first observed the optical position of the source at 8--12 GHz (X band), and detected a point source with coordinates (J2000 RA, Dec) 15:30:37.80, +26:58:56.90 and a statistical plus systematic positional uncertainty of 0.2 arcseconds in each coordinate \citep{ourAtel}. This radio emission was coincident with the optical position and had an initial flux density of 552$\pm$5\,$\mu$Jy at 9 GHz and 493$\pm$6 $\mu$Jy at 11\,GHz. We subsequently triggered follow-up spectral observations on February 27, and continued to monitor the spectral evolution of the radio emission over the following 14 months. 

The radio data were reduced in the Common Astronomy Software Application package \citep[\texttt{CASA 5.6.3,}][]{McMullin2007,CASA2022} using standard procedures, including the VLA calibration pipeline (version 5.6.3). In all observations 3C 286 was used as the flux density calibrator. 8-bit samplers were used for L- and S-band and 3-bit samplers were used for Ku-, X-, and C-bands. For phase calibration we used ICRF J151340.1+233835 for 2--18\,GHz (Ku-, X-, C-, and S-band); and ICRF J160207.2+332653 
for 1--2\,GHz (L-band). Images of the target field of view were created using the \texttt{CASA} task \texttt{tclean}, with a cell size approximately 1/5 of the synthesised beam and image sizes ranging from 1280--8000 pixels (where a larger image size was required at L-band in order to deconvolve bright sources in the field). The source flux density and associated uncertainty was measured in the image plane by fitting an elliptical Gaussian fixed to the size of the synthesized beam using the \texttt{CASA} task \texttt{imfit}, noting that a minimum uncertainty of $5\%$ of the source flux density was enforced due to the absolute flux density scale calibration accuracy of the VLA. Where enough signal-to-noise ratio
was available, we split the L-, C-, and S-band data into four sub-bands when imaging, and the X-band data into two sub-bands. The observations are summarised in Table \ref{tab:radio_obs}. 

\subsection{uGMRT}
We also observed AT2020vwl with the upgraded Giant Metrewave Radio Telescope (uGMRT) on 2021 Dec 13/14 and 2022 Apr 29. The observations were taken in band 4, with a central frequency of 0.65\,GHz and total bandwidth of 300\,MHz, and band 5, with a central frequency of 1.26\,GHz and total bandwidth of 400\,MHz. The observing bands were broken into 2048 spectral channels. Unfortunately the GWB failed for the band 5 data on 2021 Dec 13 and as a result the data were not able to be used.  
Data reduction was carried out in CASA (version 5.6.3) using standard procedures including flux and bandpass calibration with 3C286 and phase calibration with ICRF J160207.2+332653\@.
Images of the target field were again created using \texttt{tclean}. Two rounds of phase only and two rounds of phase and amplitude self-calibration were carried out on the band 4 observation. The flux density of the target was again extracted in the image plane using \texttt{imfit} by fitting an elliptical Gaussian fixed to the size and orientation of the synthesised beam. The flux densities are also listed in Table \ref{tab:radio_obs}.

\subsection{MeerKAT}
\label{sec:MeerKATobs}

Finally, we also observed AT2020vwl with the South African MeerKAT radio telescope, in the 1.3~GHz, band on
2021 August 14.7 and 2022, and in the 0.8 GHz band on 2021 Dec 27.4
and 2022 Apr 25.0 (the dates given are the midpoints of the
observations in UT).
In both bands we used the 4K (4096-channel) wideband continuum mode.
In the 1.3-GHz band, the observed bandwidth was from 856 to 1744 MHz,
with a central frequency of 1284 MHz, while in the 0.8-GHz band it was
from 544 to 1088 MHz, with a central frequency of 816 MHz.
The data were reduced using the OxKAT scripts \citep{Heywood2020_Oxkat}.  At
both bands we used ICRF J160913.3+264129 (QSO B1607+268) as a
secondary calibrator.  We used observations of ICRF J133108.2+303032
(3C~286) and ICRF J193925.0-634245 to set the flux density scale and
calibrate the bandpass at the 1.3 and 0.8 GHz bands, respectively.
The final images were made using the WSClean ($w$-stacking CLEAN)
imager \citep{Offringa+2014, OffringaS2017}, and resolved into 8
layers in frequency.  WSClean deconvolves the 8 frequency layers
together by fitting a polynomial in frequency to the brightness in the
8 frequency-layers.  Our flux densities include both the statistical
uncertainty and a systematic one due to the uncertainty in the
flux-density bootstrapping for MeerKAT, estimated at 10\% \citep[see,
  e.g.,][]{Driessen+2022}.

The flux densities were determined by fitting an elliptical Gaussian
of the same dimensions as the restoring beam to the image by least
squares.

\subsection{Interstellar scintillation}\label{sec:scintillation}

In 2022 April/May we observed AT2020vwl at $\sim$1.5\,GHz with both the VLA and uGMRT, only  9\,d apart. There was a discrepancy in flux-density between these two observations of approximately 25\%, with no corresponding discrepancy in the flux densities of background sources. Here we explore if this 
discrepancy can be explained by interstellar scintillation (ISS).

Using the NE2001 electron density model \citep{Cordes2002} we infer that for the Galactic coordinates of AT2020vwl the transition frequency between strong and weak scintillation regimes occurs at 7.4\,GHz and the angular size limit of the first Fresnel zone at the transition is 4\,micro-arcsecond. Using the \citet{Walker1998} formalism as appropriate for compact extragalactic sources, we estimate that the radio emission from AT2020vwl
will be in the strong, refractive scintillation regime until the source reaches an angular size of 134 microarcseconds. Radii of 10$^{16}$--10$^{17}$\,cm at $D_A = 138$\,Mpc would correspond to angular diameters of 10--100\,microarcesond.
The emission from AT2020vwl is expected to be affected by ISS with a timescale of variability of 67\,hr and a modulation fraction of 40$\%$ at 1.5\,GHz. We thus conclude that the 25$\%$ variation in flux density over 9\,d between the VLA and uGMRT 1.5\,GHz observations is entirely consistent with expected variability due to ISS for a source with size $<134$\,microarcseconds.

In order to account for this flux density variation, and any flux density variation due to ISS, we introduced an additional error on each radio flux density measurement. We calculated the appropriate error due to ISS for each frequency depending on the expected modulation fraction of ISS at that frequency, where the errors varied from 40$\%$ at 1.5\,GHz to 2$\%$ at 18\,GHz, and added these in quadrature with the flux density error. We report both the statistical and ISS error in Table \ref{tab:radio_obs} for each flux density measurement. In the subsequent modelling carried out below we include both the statistical and ISS errors on all flux density points. This approach resulted in  an increase in the uncertainty on the modelled parameters, however the best-fit parameter values were consistent within error of those calculated without the additional ISS error.

\subsection{Archival radio observations}

In addition to the dedicated observations described in the previous subsections, we searched for archival radio observations covering the location of AT2020vwl in order to rule out previous AGN activity in the host galaxy and to constrain any contaminating host radio  emission.

The Rapid ASKAP Continuum survey \citep[RACS,][]{McConnell2020} covered the coordinates of AT2020vwl on 2020 Oct.\ 16 at 0.88\,GHz. No source was detected at the location of AT2020vwl with a 3$\sigma$ upper limit of 837\,\muJb. 

The VLA Sky Survey \citep[VLASS][]{Lacy2020} covered the coordinates of AT2020vwl on two occasions at 3\,GHz prior to the TDE optical flare, 2017 October 02 and 2020 September 06. No source was detected in either observation with 3$\sigma$ upper limits of 342\,\muJb\ and 327\muJb\ respectively. 

The archival observations of the host galaxy thus indicate there was no significant previous AGN activity. 
However, there could be low-level host radio emission due to either a low-luminosity AGN or emission due to star formation. 

\begin{table}
	\centering
	\caption{Dedicated radio observations of AT2020vwl}
	\label{tab:radio_obs}
 \begin{tabular}{p{1.5cm}p{1.5cm}p{1cm}p{2cm}}
        \hline
        Date (UTC)& Instrument \& configuration & Frequency (GHz) & Flux Density $\pm$statistical error $\pm$ISS error ($\mu$Jy)\\
        \hline
2021-02-03 & VLA-A & 9.0  & 552$\pm$28$\pm$11.04 \\
 &  & 11.0  & 493$\pm$25$\pm$9.86 \\
 \hline
2021-02-07 & VLA-A & 1.26  & 244$\pm$46$\pm$97.6 \\
 &  & 1.78  & 294$\pm$33$\pm$117.6 \\
 &  & 3.0  & 412$\pm$21$\pm$123.6 \\
 &  & 4.5  & 500$\pm$25$\pm$100. \\
 &  & 5.51  & 543$\pm$27$\pm$108.6 \\
 &  & 6.49  & 576$\pm$29$\pm$115.2 \\
 &  & 7.51  & 564$\pm$28$\pm$11.28 \\
 &  & 9.0  & 517$\pm$28$\pm$10.34 \\
 &  & 11.0  & 565$\pm$28$\pm$11.3 \\
 \hline
2021-05-07& VLA-D & 2.75  & 461$\pm$53$\pm$138.3 \\
 &  & 3.25  & 476$\pm$53$\pm$142.8 \\
 &  & 5.51  & 504$\pm$25$\pm$100.8 \\
 &  & 6.49  & 495$\pm$24$\pm$99 \\
 &  & 7.51  & 470$\pm$24$\pm$9.4 \\
 &  & 9.0  & 440$\pm$22$\pm$8.8 \\
 &  & 11.0  & 390$\pm$20$\pm$7.8 \\
 &  & 15.0  & 297$\pm$15$\pm$5.94 \\
 \hline
2021-08-14 & MeerKAT & 1.284  & 373$\pm$40$\pm$146.8 \\
2021-08-11 & VLA-C & 1.52  & 366$\pm$70$\pm$146.4 \\
 &  & 2.5  & 457$\pm$37$\pm$137.1 \\
 &  & 3.24  & 468$\pm$30$\pm$140.4 \\
 &  & 4.49  & 410$\pm$34$\pm$82 \\
 &  & 5.51  & 284$\pm$23$\pm$56.8 \\
 &  & 6.49  & 266$\pm$23$\pm$53.2 \\
 &  & 7.51  & 207$\pm$20$\pm$4.14 \\
 &  & 9.0  & 186$\pm$9$\pm$3.72 \\
 &  & 11.0  & 147$\pm$9$\pm$2.94 \\
 &  & 15.08  & 115$\pm$6$\pm$2.3 \\
 \hline
2021-10-18 & VLA-B  & 1.26  & 355$\pm$39$\pm$142 \\
 &  & 1.65  & 490$\pm$92$\pm$196 \\
 &  & 1.9  & 353$\pm$78$\pm$141.2 \\
 &  & 2.24  & 301$\pm$54$\pm$90.3 \\
 &  & 2.75  & 274$\pm$29$\pm$82.2 \\
 &  & 3.24  & 247$\pm$25$\pm$74.1 \\
 &  & 3.75  & 259$\pm$25$\pm$77.7 \\
 &  & 5.61  & 156$\pm$25$\pm$31.2 \\
 &  & 6.61  & 160$\pm$22$\pm$32 \\
 &  & 7.57  & 135$\pm$22$\pm$2.7 \\
 &  & 9.0  & 109$\pm$13$\pm$2.18 \\
 &  & 15.08  & 53$\pm$6$\pm$1.054 \\
 \hline
2021-12-14 & uGMRT & 0.65  & 440.0$\pm$66.0$\pm$110 \\
2021-12-27 & MeerKAT & 0.815  & 348.0$\pm$47.0$\pm$87 \\
2021-12-14 & VLA-B &1.25  & 381$\pm$36$\pm$152.4 \\
 &  & 1.75  & 332$\pm$28$\pm$132.8 \\
 &  & 2.31  & 280$\pm$42$\pm$84 \\
 &  & 2.87  & 228$\pm$26$\pm$68.4 \\
 &  & 3.3  & 227$\pm$25$\pm$68.1 \\
 &  & 3.75  & 134$\pm$27$\pm$40.2 \\
 &  & 5.0  & 145$\pm$15$\pm$29 \\
 &  & 7.0  & 109$\pm$13$\pm$13 \\
 &  & 9.0  & 81$\pm$9$\pm$1.62 \\
 &  & 11.0  & 70$\pm$11$\pm$1.4 \\
 \hline
 \end{tabular}
\end{table}
\begin{table}
	\centering
	\caption{Table \ref{tab:radio_obs} continued}
	\label{tab:radio_obs2}
 \begin{tabular}{p{1.5cm}p{1.5cm}p{1cm}p{2cm}}
        \hline
        Date (UTC)& Instrument & Frequency (GHz) & Flux Density $\pm$statistical error $\pm$ISS error  ($\mu$Jy)\\
        \hline
2022-04-29 & uGMRT & 0.65  & 521.0$\pm$217.3$\pm$130.25 \\
2022-04-24 & MeerKAT & 0.81593  & 671.0$\pm$65.0$\pm$167.75 \\
2022-04-29 & uGMRT & 1.26  & 668.0$\pm$253.4$\pm$267.2 \\
2022-05-11 & MeerKAT & 1.284  & 356$\pm$37$\pm$184.4 \\
2022-05-08 & VLA-A  & 1.5  & 406$\pm$19$\pm$162.4 \\
 &  & 2.243  & 294$\pm$37$\pm$88.2 \\
 &  & 2.754  & 298$\pm$21$\pm$89.4 \\
 &  & 3.24  & 208$\pm$17$\pm$62.4 \\
 &  & 3.75  & 225$\pm$17$\pm$67.5 \\
 &  & 4.48  & 235$\pm$18$\pm$47 \\
 &  & 5.51  & 179$\pm$17$\pm$35.8 \\
 &  & 6.49  & 141$\pm$15$\pm$28.2 \\
 &  & 7.45  & 120$\pm$14$\pm$2.4 \\
 &  & 10.0  & 68$\pm$6$\pm$1.36 \\

  \hline
  \end{tabular}
\end{table}

\subsection{Archival optical observations of the host galaxy}\label{sec:optical}

In order to disentangle the transient radio emission due to the TDE from any emission intrinsic to the host galaxy, we analysed the host galaxy properties based on publicly available data. 

\begin{figure}
    \centering
	\includegraphics[width=0.7\columnwidth]{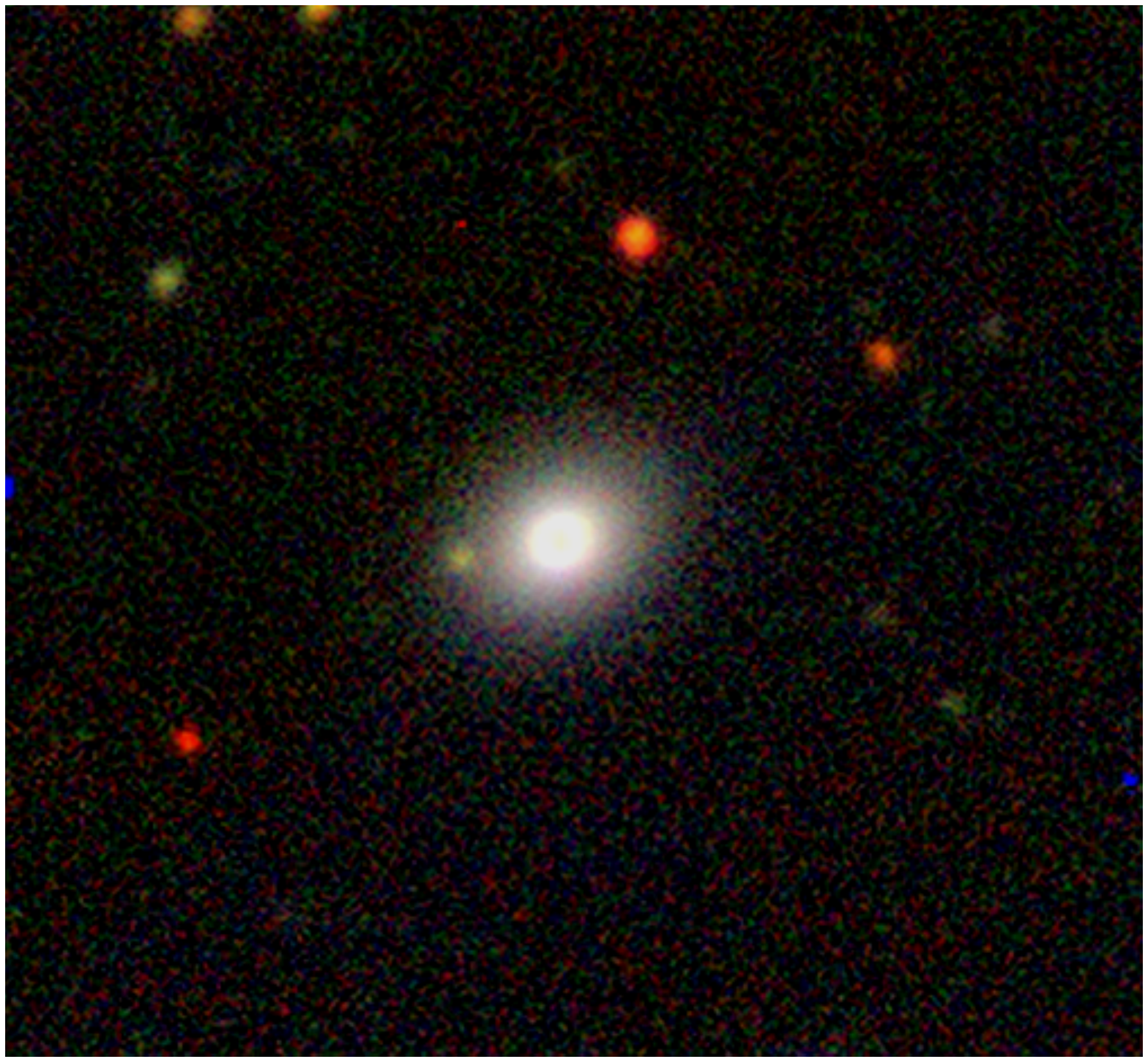}
    \caption{DeCals \citep{Dey2019} image of SDSS J153037.80+265856.8, the host galaxy of AT2020vwl. The galaxy appears to be lenticular with little visual evidence of spiral structure, indicating it is unlikely to be an active star-forming galaxy.}
    \label{fig:host_image}
\end{figure}

\begin{figure*}
	\includegraphics[width=2\columnwidth]{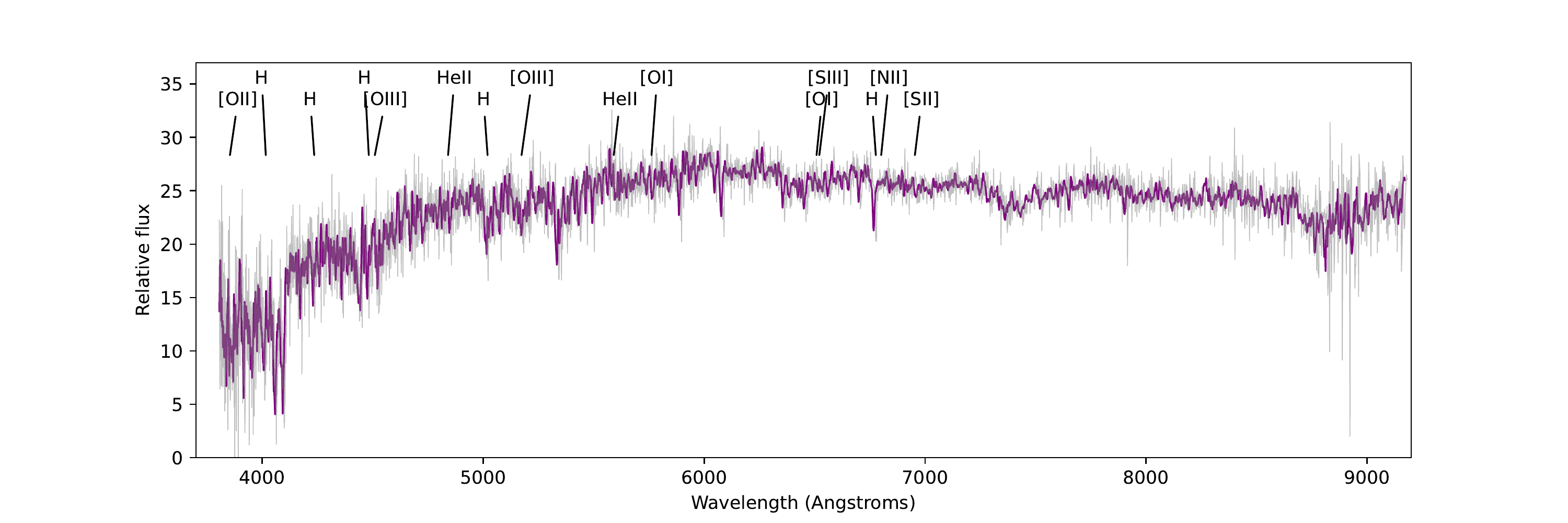}
    \caption{SDSS optical spectrum of SDSS J153037.80+265856.8, the host galaxy of AT2020vwl, taken on MJD 54180 (13.5 years before the tidal disruption event occurred). The spectrum shows a quiescent galaxy with no obvious AGN or active star formation, indicating that the galaxy is unlikely to have significant radio emission due to AGN or star formation activity.}
    \label{fig:sdss_spec}
\end{figure*}

The host galaxy of AT2020vwl, SDSS J153037.80+265856.8/LEDA 1794348, appears to be lenticular, with little evidence of spiral structure (Figure~\ref{fig:host_image}). An  SDSS \citep[Sloan Digital Sky Survey][]{York2000} spectrum \citep{Strauss2002} of the host galaxy taken 13.5\,yr prior to the TDE on MJD 54180 shows a quiescent galaxy spectrum with no sign of strong AGN emission lines (e.g.\ [OIII] or [NII]), Figure~\ref{fig:sdss_spec}). There is no evidence for H$\alpha$ emission in the spectrum, which would be indicative of shocked or excited gas due to either AGN activity or active star formation in the galaxy. 
The WISE \citep[Wide-field Infrared Survey Explorer;][]{Wright2010} photometry for the host galaxy (WISEA J153037.80+265856.8) at the different WISE filters are as follows: W1=13.989$\pm$0.025 and W2=14.001$\pm$0.037. The WISE W1-W2 color of $-0.012\pm0.03$ also does not indicate any AGN or star-forming activity (cf.\ W1-W2$>$0.8 would indicate AGN activity \citep{Stern2012}).  

These spectral properties, combined with the lack of a radio detection of the host prior to the TDE, indicate that the galaxy likely does not host an AGN and is not an active star forming galaxy. Therefore, most of the radio emission observed is likely intrinsic to the transient event. 

The optical spectral properties of the host enable an approximate estimate of the star formation rate of 0.002--0.22 \Msolxyr\ \citep[][using the FSPS-Granada catalogue given reasonable model variations of dust and IMF]{Conroy2009}. Using the \citet{Murphy2011} star formation relation at 1.4\,GHz, we can infer the 1.4\,GHz luminosity expected for this star formation rate of the galaxy,

\begin{align}
    \left(\frac{\rm{SFR}_{1.4\,\rm{GHz}}}{\Msolxyr}\right) = 6.35 \times 10^{-29} \left(\frac{L_{\rm{1.4 \,GHz}}}{\rm{erg\,s^{-1}\,Hz^{-1}}}\right),
\end{align}
where $L_{\rm{1.4 GHz}}$ is the 1.4\,GHz luminosity of the galaxy. \citet{Murphy2011} found that this relation had a residual dispersion of $\sigma$=0.3--0.5 when comparing various extinction independent SFR diagnostics. The expected scatter in the relation is therefore significantly smaller than the uncertainty of the measured SFR of the host galaxy.

Such a low star formation rate would be expected to give rise to a small amount of radio emission, with $L_{\rm 1.4\,GHz}=3.1\times10^{25} - 3.4\times10^{27}$\,erg s$^{-1}$ Hz$^{-1}$, or a flux density of 1.4--149\,$\mu$Jy at the distance of AT2020vwl. Thus we can conclude that the star formation contribution to the radio emission observed from the host galaxy may not be negligible, motivating the use of a host component in the transient modelling outlined below. 

\subsection{Archival \textit{Swift} observations}

We searched the Neil Gehrels Swift Observatory (\textit{Swift}) \citep{Burrows2005} archive for publicly available observations of AT2020vwl. Between 2021 Jan 07 and 2022 Dec 05 there were 27 observations of the source taken with the \textit{Swift} X-ray Telescope (XRT) and Ultra-violet Optical Telescope (UVOT). To search for any X-ray counterpart to the event, we examined all 27 observations using the \textit{Swift} online XRT product builder \citep{Evans2009}. The XRT observations were taken in photon counting (PC) mode. In all observations there was no X-ray source detected at the position of AT2020vwl, with a 3$\sigma$ upper limit on the 0.2-10\,keV X-ray flux on 2021 Jan 07 of $F_X < 4.2\times10^{-12}$\,erg\,cm$^{-2}$\,s$^{-1}$ (assuming a distance of $z=0.035$, Galactic hydrogen column density of $N_H=4.3\times10^{20}$\,cm$^{-2}$ \citep{Willingale2013} and photon index $\Gamma=1.5$). 

UVOT observations included measurements with the UVW1 (peak sensitivity at 2600$\dot{A}$), UVW2 (peak sensitivity at 1928$\dot{A}$), V (peak sensitivity at 5468$\dot{A}$), U (peak sensitivity at 3465$\dot{A}$), UVM2 (peak sensitivity at 2245$\dot{A}$) and B (peak sensitivity at 4392$\dot{A}$) filters. We extracted the UV flux of AT2020vwl as measured by UVOT, using the HEAsoft Swift software tools\footnote{\url{https://heasrc.nasa.gov/lheasoft/}} \texttt{UVOTSOURCE} task to carry out aperture photometry and extract a UV lightcurve. We used a circular source region of 5" and background region consisting of a 20" aperture nearby to the source. We extinction corrected the UV magnitudes using extinction estimates derived from \citet{Schlegel1998}. The error bars correspond to 1$\sigma$. AT2020vwl was detected in all UVOT observations, with U magnitudes plotted in Figure \ref{fig:optical_lc}.

\section{Results}\label{sec:results}

\begin{figure}
	\includegraphics[width=\columnwidth]{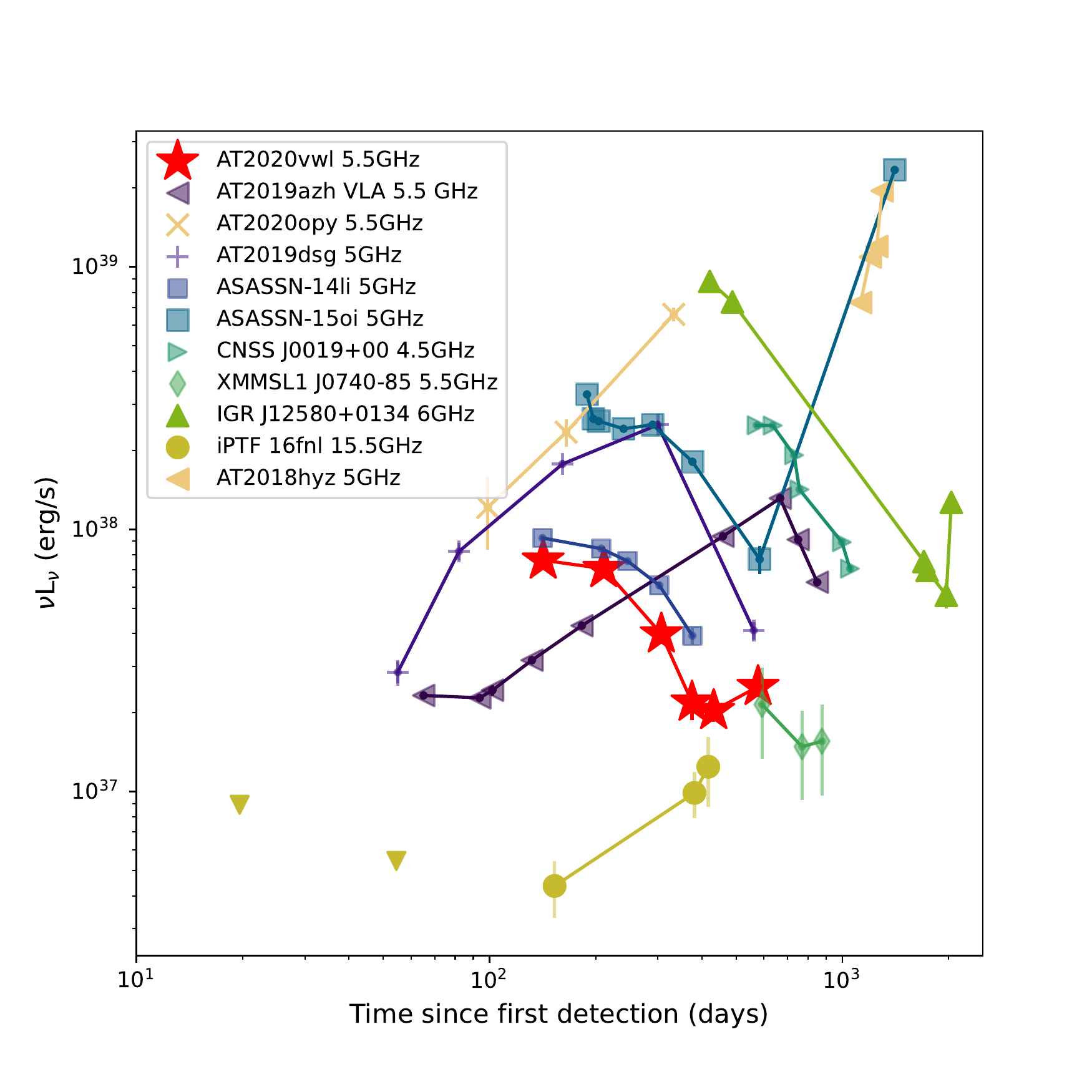}
    \caption{The 5.5\,GHz radio luminosity curve of AT2020vwl (red stars) compared to those of selected other radio-bright thermal TDEs. TDE data are from AT2020opy \citet{Goodwin2023}; AT2019azh \citet{Goodwin2022}; AT2019dsg \citet{Cendes2021}; ASASSN-14li \citet{Alexander2016}; ASASSN-15oi \citet{Horesh2021}; AT2018hyz \citet{Cendes2022}; CNSS J0019+00 \citet{Anderson2020}; XMMSL1 J0740-85 \citet{Alexander2017}; IGR J12850+0134 \citet{Perlman2022}, \citet{Lei2016}, and \citet{Nikolajuk2013}; iPTF 16fnl \citet{Horesh2021b}. The x-axis indicates the time since the first detection (optical, radio, or X-ray depending on the source) of each TDE.}
    \label{fig:lightcurve}
\end{figure}

\begin{figure}
	\includegraphics[width=\columnwidth]{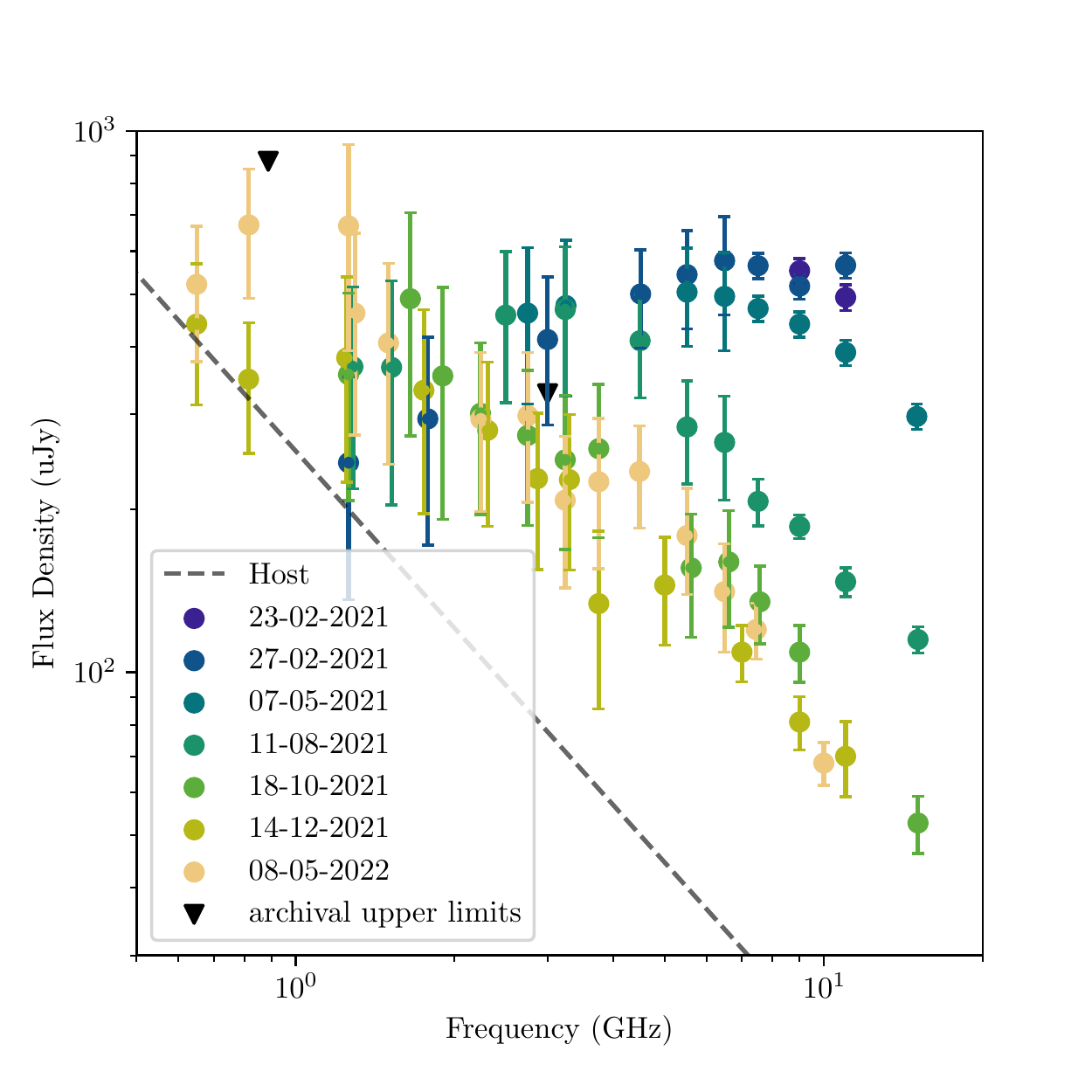}
    \caption{Radio emission from AT2020vwl, observed  with the VLA, uGMRT,
    and MeerKAT radio telescopes. Archival 3-$\sigma$ upper limits from VLASS (3\,GHz) and RACs (0.88\,GHz) are shown in inverted triangles. The assumed underlying host component of the emission is shown in dashed grey. }
    \label{fig:rawradio}
\end{figure}

We show the 5.5\,GHz lightcurve for AT2020vwl, as well as a comparison to other TDEs in Figure \ref{fig:lightcurve}. 
The radio emission from AT2020vwl gradually faded over the course of our radio observations from 142--432\,d post-optical detection, evolving on a timescale similar to that of the thermal TDE ASASSN-14li \citep{Alexander2016}. There is a slight increase in the overall flux density in the final epoch at 577\,d. which could be due to an increase in the energy in the outflow. 
The individual radio spectra over 0.65--15\,GHz for each epoch are shown in Figure \ref{fig:rawradio}. The radio emission shows a peaked spectrum, initially peaking at $\approx$7\,GHz with the peak shifting to lower frequencies over our observing period.

\subsection{Optical lightcurve}

\begin{figure}
    \centering
    \includegraphics[width=\columnwidth]{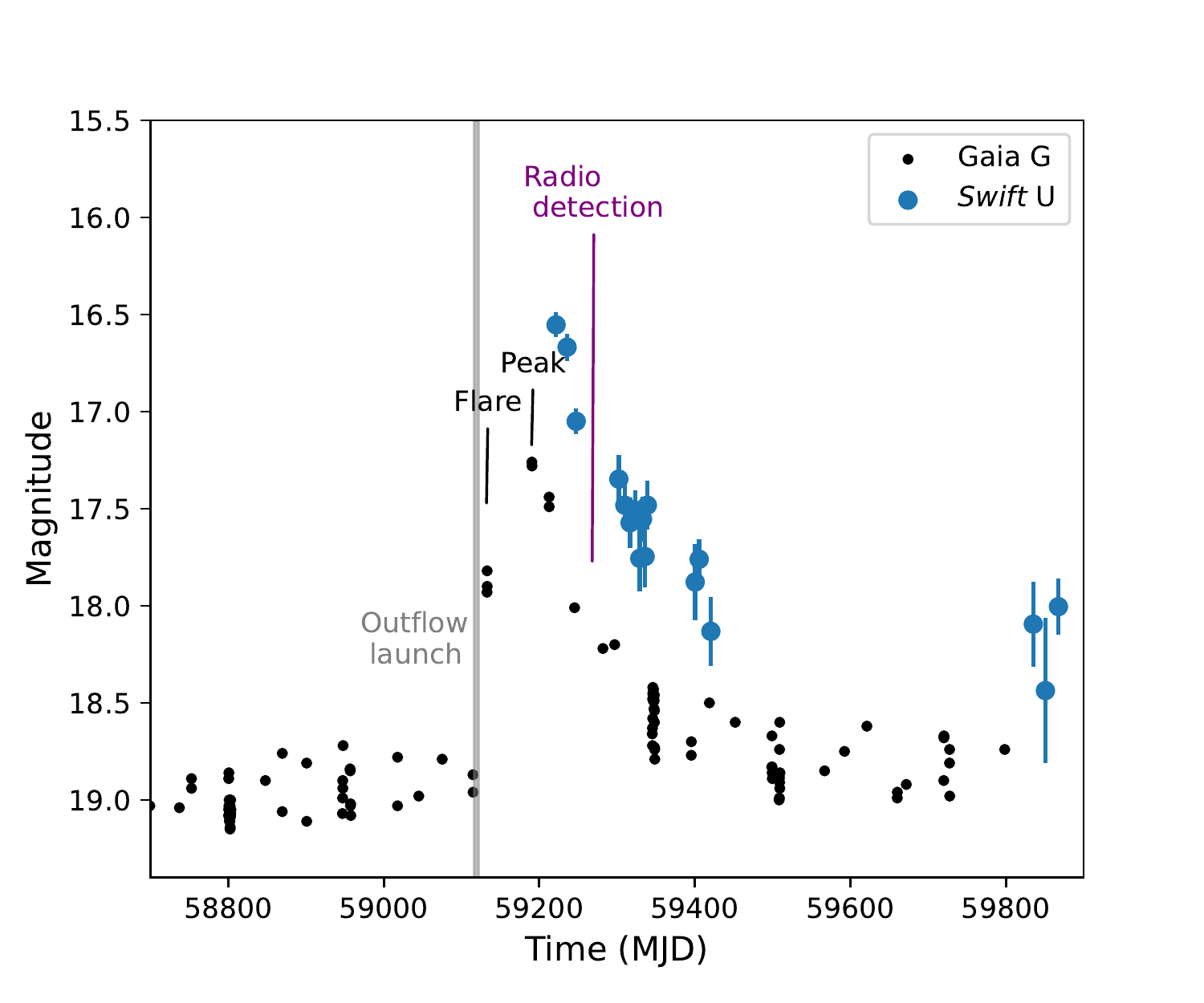}
    \caption{Gaia G-band (black) and \textit{Swift} U-band (blue) lightcurves of AT2020vwl (Gaia20etp). The initial optical detection on 2020 October 10, peak optical flux on 2020 December 07, and the initial radio detection on 2021 February 23 are indicated. The inferred radio outflow launch time is indicated in grey, where the shaded region denotes the 1$\sigma$ uncertainty on the outflow launch date.
    }
    \label{fig:optical_lc}
\end{figure}

In Figure \ref{fig:optical_lc} we plot the optical lightcurve of LEDA 1794348 from the \textit{Gaia} Spacecraft \citep{Gaia2016}, which reported 
transient optical activity of the TDE AT2020vwl \citep{Gaia_ATel}. The optical flare of $\sim1$\,mag first occurred on 2020 October 10 (MJD 59132), with a previous detection of the host galaxy 18 days earlier on 2020 September 22 (MJD 59114). The Gaia lightcurve is very well sampled, and shows the optical emission rose slowly to a peak on approximately 2020 Dec 07 (MJD 59190), 58\,d later, consistent with average rise times of other TDEs \citep{vanVelzen2021}. 

\subsection{Radio spectral fitting}\label{sec:specfits}

We fit the observed radio spectra for each epoch with a synchrotron emission model. While the host galaxy optical properties indicate that we do not expect significant host radio emission due to AGN activity, the inferred star formation rate from archival optical spectra implies there could be a small amount of radio flux unrelated to the TDE. Therefore, our synchrotron emission model consists of two components:  the first is a broken powerlaw, representing a component which is synchrotron self-absorbed at low frequencies, and the second is an unbroken powerlaw, representing the host
galaxy emission.  Such a two-part model is described in  \citet{Alexander2016} and \citet{Goodwin2022}. In this model, the flux density of the self-absorbed synchrotron component is given by \citet{Granot2002} 

\begin{equation}
    \label{eq:Fv}
    \begin{aligned}
        F_{\nu, \mathrm{synch}} = F_{\nu,\mathrm{ext}} \left[\left(\frac{\nu}{\nu_{\rm m}}\right)^2 \exp(-s_1\left(\frac{\nu}{\nu_{\rm m}}\right)^{2/3}) + \left(\frac{\nu}{\nu_{\rm m}}\right)^{5/2}\right] \times \\
        \left[1 + \left(\frac{\nu}{\nu_{\rm a}}\right)^{s_2(\beta_1 - \beta_2)}\right]^{-1/s_2}
        \end{aligned}
    \end{equation}
    where $\nu$ is the frequency, $F_{\nu,\mathrm{ext}}$ is the normalisation, $s_1 = 3.63p-1.60$, $s_2 = 1.25-0.18p$, $\beta_1 = \frac{5}{2}$, $\beta_2 = \frac{1-p}{2}$, and $p$ is the energy index of the powerlaw distribution of relativistic electrons, $\nu_{\rm m}$ is the synchrotron minimum frequency, and $\nu_{\rm a}$ is the synchrotron self-absorption frequency. 
    We assume further that $\nu_{\rm m} < \nu_{\rm a} < \nu_{\rm c}$,  where $\nu_{\rm c}$ is the synchrotron cooling frequency. 

The flux density of the host component is 
\begin{equation}
    \label{eq:Fhost}
    F_{\nu,\rm{host}} = F_0 \left(\frac{\nu}{1.4\,\rm{GHz}}\right)^{\alpha_0}
\end{equation}
where $F_0$ is the flux density measured at 1.4\,GHz and $\alpha_0$ is the spectral index of the host galaxy. 

The total observed flux density model is then

\begin{equation}
    F_{\nu, \rm{total}} =  F_{\nu,\rm{host}} + F_{\nu, \mathrm{synch}}
\end{equation}

In order to constrain $F_0$ and $\alpha_0$, we fit three of the most well-constrained spectra (2021-05-07, 2021-08-11, and 2021-10-18) as outlined in the next paragraph, but also including $F_0$ and $\alpha_0$ as parameters in the fit. Due to archival observations of the host galaxy (see Section \ref{sec:observations}), we constrain $F_0$ to $<0.5$\,mJy.  
The values of $F_0$ and $\alpha$ obtained agreed within the uncertainties.
We adopt the mean values of  $F_0=0.178\pm0.05$\,mJy and $\alpha_0=-1.1\pm0.2$ from these three fits for our other fits. The host galaxy component corresponding to these values is plotted in Figure~\ref{fig:rawradio}. 
This flux density corresponds to a 1.4~GHz $L_\nu$ of
$L_{1.4\rm{GHz}}=4.6\times10^{27}$\,erg\,s$^{-1}$\,Hz$^{-1}$, which is
consistent with the optical properties of the host showing a lack of significant AGN or star formation activity, and consistent with the flux density estimate of the radio emission due to a small amount of ongoing star formation in the galaxy of $L_{\rm 1.4\,GHz}=3.1\times10^{25}-3.4\times10^{27}$\,erg s$^{-1}$ Hz$^{-1}$ (Section \ref{sec:optical}).
We note that the assumed host emission is accounting for possible low-luminosity AGN activity, as well as the approximate expected radio emission from star formation in the galaxy.

Additionally, in Appendix \ref{sec:Appendix} we present a statistical comparison of the spectral fits both with and without the host emission, and find that the first epoch is better fit with the model including a host component while the other epochs are equally well-fit by the model with or without the host component. In the first epoch the peak of the transient emission has not yet evolved to lower frequencies where the host emission dominates, which could explain the preference for host emission in the first epoch but not the latter epochs.  

We fit the flux density as a function of frequency for each epoch using a Python implementation of Markov Chain Monte Carlo (MCMC), \texttt{emcee} \citep{emcee} and Equation \ref{eq:Fv}, with, as mentioned, a fixed contribution from the host
galaxy contribution as given by Eq.~\ref{eq:Fhost}.  This approach enables us to obtain posterior distributions for the $p, F_{\nu,\mathrm{ext}},
\nu_{\rm m}$ and $\nu_{\rm a}$.
We assume flat prior distributions for all parameters, constraining $p$ to the range 2.5--4.0, $F_{\nu,\rm{ext}}$ to the range $10^{-6}$--$1$, $\nu_m$ to the range 0.5--$\nu_a$, and $\nu_a$ to the range $\nu_b$--12.  For each parameter we report the median value from the posterior distribution and the 16th and 84th percentiles, corresponding to approximately 1$\sigma$ errors. For each MCMC calculation we use 400 walkers and 4000 steps, discarding the first 1000 steps to account for burn-in. 

\begin{figure}
	\includegraphics[width=\columnwidth]{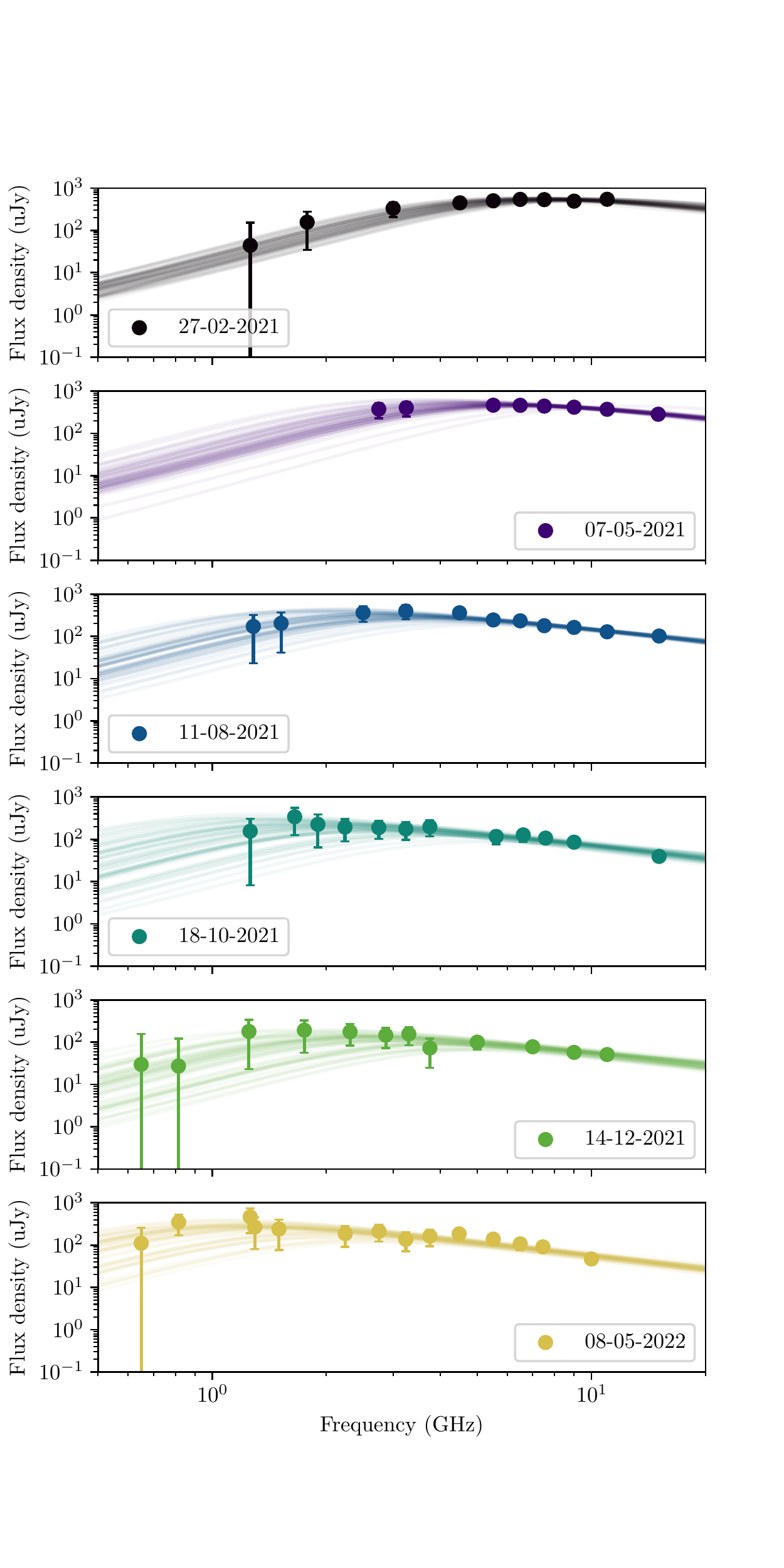}
    \caption{Synchrotron spectral fits of the evolving radio emission observed from AT2020vwl between 2020 February to 2022 May. The observed radio flux densities minus the assumed host component are plotted in circles and 50 random MCMC samples are plotted in solid lines to demonstrate the uncertainty in the fits. The radio emission is well described by a synchrotron self-absorption spectrum that evolves to peaking at lower frequencies over time.}
    \label{fig:spectrum}
\end{figure}

We plot the spectral fits for each epoch in Figure \ref{fig:spectrum}, and report the best fit flux densities and frequencies of the spectral peaks for each epoch
in Table \ref{tab:specproperties}. Both the flux density and frequency of the spectral peaks consistently decreased over the course of our observations, whilst $p$ remained approximately constant at $p\approx3$. 

\begin{table}
	\centering
	\caption{Synchrotron emission modelling properties of the outflow produced by the TDE AT2020vwl}
	\label{tab:specproperties}
 \begin{tabular}{p{1.35cm}p{0.85cm}p{1cm}p{1.cm}p{1.cm}p{1cm}}
Date (UTC) & $\delta$t (d)$^{a}$ & $F_{\rm{p}}$ (mJy) & $\nu_{\rm{m}}$ (GHz) & $\nu_{\rm{p}}$ (GHz) & $p$ \\
\hline
27-02-2021 & 142 &
0.59$\pm$0.03& 3.69$\pm$1.00 & 8.04$\pm$0.81 & 2.85$\pm$0.27\\
07-05-2021 & 211 &
0.54$\pm$0.05& 2.56$\pm$0.81 & 5.79$\pm$0.68 & 2.89$\pm$0.26\\
11-08-2021 & 307 &
0.35$\pm$0.06& 1.50$\pm$0.56 & 3.05$\pm$0.53 & 2.96$\pm$0.24\\
18-10-2021 & 375 &
0.21$\pm$0.06& 1.25$\pm$0.51 & 2.35$\pm$0.54 & 3.07$\pm$0.22\\
14-12-2021 & 432 &
0.13$\pm$0.03& 1.48$\pm$0.73 & 2.90$\pm$0.81 & 2.95$\pm$0.26\\
08-05-2022 & 577 &
0.26$\pm$0.07& 0.87$\pm$0.31 & 1.55$\pm$0.36 & 3.09$\pm$0.21\\
\hline
\hline
\end{tabular}
\\
\footnotesize{$^{a}$ $\delta$t is measured with respect to the initial optical detection, $t_0=$ MJD 59130}
\end{table}

\begin{table*}
	\centering
	\caption{Equipartition modelling properties of the outflow produced by the TDE AT2020vwl. We report both the uncorrected equipartition radius ($R_{\rm{eq}}$) and energy ($E_{\rm{eq}}$) as well as the corrected radius ($R$) and energy ($E$) assuming 2$\%$ of the energy is carried by the magnetic field and 10$\%$ of the energy is carried by the electrons. }
	\label{tab:outflowproperties}
 \begin{tabular}{p{1.5cm}p{0.85cm}p{1.5cm}p{1.5cm}p{1.3cm}p{1.3cm}p{1.3cm}p{1.6cm}p{1.3cm}p{1.6cm}}
        \hline
&$\delta$t (d)$^{a}$& log $R_{\rm{eq}}$ (cm) &  log $E_{\rm{eq}}$ (erg) & log $E$ (erg) &  log $R$ (cm) &
 $\beta$ (c) & log $n_{e}$ (cm$^{-3}$) & log $B$ (G) & log $M_{\rm ej}$ (M$_{\odot}$) \\
\hline

Spherical, & 142 & 16.02$\pm$0.06 & 47.74$\pm$0.06 &
47.99$\pm$0.06 & 16.08$\pm$0.06 &    0.032$\pm$0.004 &     3.83$\pm$0.46 &     -0.12$\pm$0.35 &     -2.99$\pm$0.08\\
$\epsilon_e=0.1$& 211 & 16.15$\pm$0.06 & 47.86$\pm$0.08 &
48.11$\pm$0.08 & 16.20$\pm$0.06 &    0.029$\pm$0.004 &     3.58$\pm$0.53 &     -0.25$\pm$0.40 &     -2.78$\pm$0.10\\
& 307 & 16.34$\pm$0.09 & 47.97$\pm$0.12 &
48.22$\pm$0.12 & 16.40$\pm$0.09 &    0.032$\pm$0.006 &     3.11$\pm$0.75 &     -0.48$\pm$0.55 &     -2.73$\pm$0.15\\
& 375 &16.36$\pm$0.11 & 47.91$\pm$0.17 &
48.16$\pm$0.17 & 16.42$\pm$0.11 &    0.027$\pm$0.007 &     2.99$\pm$0.97 &     -0.55$\pm$0.68 &     -2.66$\pm$0.20\\
& 432 & 16.16$\pm$0.12 & 47.47$\pm$0.17 &
47.72$\pm$0.17 & 16.22$\pm$0.12 &    0.015$\pm$0.004 &     3.14$\pm$1.04 &     -0.47$\pm$0.76 &     -2.59$\pm$0.21\\
& 577 & 16.58$\pm$0.10 & 48.20$\pm$0.16 &
48.45$\pm$0.16 & 16.64$\pm$0.10 &    0.029$\pm$0.007 &     2.63$\pm$0.91 &     -0.73$\pm$0.63 &     -2.43$\pm$0.19\\
\hline
Conical, & 142 &16.40$\pm$0.06 & 48.26$\pm$0.06 &48.51$\pm$0.06 & 16.45$\pm$0.06 &    0.074$\pm$0.009 &     3.29$\pm$0.46 &     -0.39$\pm$0.35 &     -3.18$\pm$0.08\\
$\epsilon_e=0.1$& 211 &16.52$\pm$0.06 & 48.39$\pm$0.08 &48.64$\pm$0.08 & 16.58$\pm$0.06 &    0.067$\pm$0.010 &     3.04$\pm$0.53 &     -0.52$\pm$0.40 &     -2.97$\pm$0.10\\
& 307 & 16.72$\pm$0.09 & 48.50$\pm$0.12 &48.75$\pm$0.12 & 16.77$\pm$0.09 &    0.072$\pm$0.014 &     2.58$\pm$0.75 &     -0.75$\pm$0.55 &     -2.91$\pm$0.15\\
& 375 &16.72$\pm$0.11 & 48.45$\pm$0.17 &48.70$\pm$0.17 & 16.79$\pm$0.11 &    0.062$\pm$0.016 &     2.47$\pm$0.97 &     -0.81$\pm$0.68 &     -2.84$\pm$0.20\\
& 432 & 16.54$\pm$0.12 & 48.00$\pm$0.17 &48.25$\pm$0.17 & 16.60$\pm$0.12 &    0.035$\pm$0.010 &     2.61$\pm$1.04 &     -0.74$\pm$0.76 &     -2.79$\pm$0.21\\
& 577 & 16.96$\pm$0.10 & 48.74$\pm$0.16 &48.99$\pm$0.16 & 17.01$\pm$0.10 &    0.067$\pm$0.016 &     2.10$\pm$0.91 &     -0.99$\pm$0.63 &     -2.61$\pm$0.19\\

\hline
Spherical, & 142 &&&
50.04$\pm$0.06 & 16.33$\pm$0.06 &    0.057$\pm$0.007 &     1.09$\pm$0.47 &     0.90$\pm$0.36 &     -1.42$\pm$0.08\\
$\epsilon_e=5\times10^{-4}$& 211 & & &
50.16$\pm$0.07 & 16.46$\pm$0.06 &    0.052$\pm$0.007 &     0.82$\pm$0.53 &     0.77$\pm$0.40 &     -1.22$\pm$0.10\\
& 307 & & &
50.26$\pm$0.12 & 16.65$\pm$0.09 &    0.055$\pm$0.011 &     0.33$\pm$0.77 &     0.53$\pm$0.56 &     -1.18$\pm$0.15\\
& 375 & & &
50.18$\pm$0.16 & 16.67$\pm$0.11 &    0.048$\pm$0.012 &     0.17$\pm$0.96 &     0.47$\pm$0.68 &     -1.13$\pm$0.20\\
& 432 & & &
49.75$\pm$0.18 & 16.47$\pm$0.12 &    0.027$\pm$0.008 &     0.37$\pm$1.07 &     0.55$\pm$0.77 &     -1.05$\pm$0.22\\
& 577 & & &
50.47$\pm$0.16 & 16.89$\pm$0.10 &    0.051$\pm$0.012 &     -0.20$\pm$0.90 &     0.28$\pm$0.63 &     -0.90$\pm$0.19\\
\hline
Conical, & 142 & & &50.56$\pm$0.06 & 16.71$\pm$0.06 &    0.126$\pm$0.016 &     0.55$\pm$0.47 &     0.63$\pm$0.36 &     -1.59$\pm$0.08\\
$\epsilon_e=5\times10^{-4}$& 211 & & &50.68$\pm$0.07 & 16.83$\pm$0.06 &    0.115$\pm$0.017 &     0.28$\pm$0.53 &     0.51$\pm$0.40 &     -1.39$\pm$0.10 \\
& 307 & & &50.79$\pm$0.12 & 17.03$\pm$0.09 &    0.122$\pm$0.025 &     -0.21$\pm$0.77 &     0.27$\pm$0.56 &     -1.34$\pm$0.15\\
& 375 & & &50.72$\pm$0.16 & 17.05$\pm$0.11 &    0.106$\pm$0.026 &     -0.36$\pm$0.96 &     0.20$\pm$0.68 &     -1.29$\pm$0.20\\
& 432 & & &50.28$\pm$0.18 & 16.85$\pm$0.12 &    0.061$\pm$0.017 &     -0.17$\pm$1.07 &     0.28$\pm$0.77 &     -1.25$\pm$0.22\\
& 577 & & &51.01$\pm$0.16 & 17.27$\pm$0.10 &    0.114$\pm$0.026 &     -0.73$\pm$0.90 &     0.02$\pm$0.63 &     -1.05$\pm$0.19\\

\hline
\hline
	\end{tabular}
 \\
	\footnotesize{$^{a}$ $\delta$t is measured with respect to the initial optical detection, $t_0=$ MJD 59130}
\end{table*}

\subsection{Outflow modelling}\label{sec:modelling}

In order to estimate the physical properties of the outflow, we first assume equipartition between the energies in the relativistic electrons and the magnetic field, and infer the outflow properties by assuming the outflow takes the form of a blastwave that accelerates ambient electrons into a power-law distribution, $N(\gamma)\propto \gamma^{-p}$, where $p$ is the synchrotron energy index,
$\gamma$ is the electron Lorentz factor, 
with $\gamma\geq \gamma_{\rm m}$ where $\gamma_{\rm m}$ is the minimum Lorentz factor. We use the approach outlined by \citet{BarniolDuran2013} to estimate key physical quantities such as the radius ($R$) and energy ($E$) of the outflow, the magnetic field strength ($B$), mass of the emitting region ($M_{\rm ej}$), ambient electron density ($n_e$, calculated based on the inferred total number of electrons in the observed region, $N_e$), and velocity of the ejecta ($\beta$). The exact equations we used are Equations 4--13 in \citet{Goodwin2022}. 

\begin{figure*}
	\includegraphics[width=\textwidth]{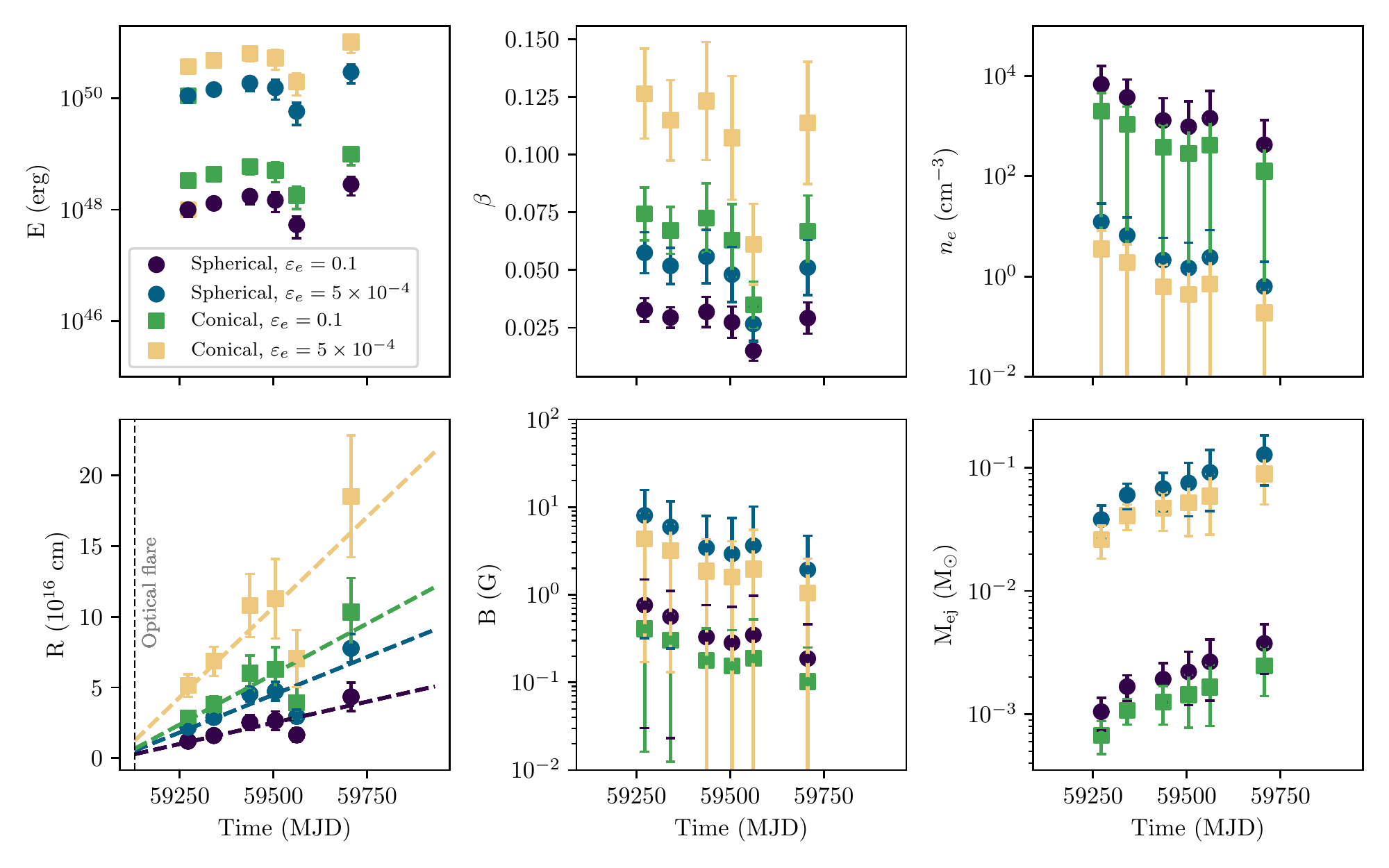}
    \caption{Physical outflow properties inferred by equipartition modelling of the spectral properties of the radio emission from the TDE AT2020vwl.
    Properties derived with an assumed spherical geometry are plotted in \textbf{circles}, and ones with a conical geometry in \textbf{squares}.$E$ and $R$ are the estimated energy and radius of the outflow derived from an equipartition analysis and corrected for assumed deviation from equipartition. Two assumptions about the deviation from equipartition are shown, one where the fraction of energy in the electrons is $\epsilon_e=0.1$ (purple and green) and one where $\epsilon_e=5\times10^{-4}$ (blue and yellow).}  $B$ is the magnetic field strength, $M_{\rm ej}$ is the mass in the ejecta, $\beta$ is the outflow velocity divided by the speed of light, and $n_e$ is the free electron number density of the ambient medium. The dashed lines in the lower left panel show a linear fit to the radius for each geometry. The energy and radius increased with time until 307\,d post-disruption, at which time the radius plateaued and the energy decreased until a renewed increase in radius and energy was observed at 577\,d.
    \label{fig:outflowmodelling}
\end{figure*}

As in \citet{Goodwin2022}, we first assume equipartition to derive the equipartition radius ($R_{\rm{eq}}$) and energy ($E_{\rm{eq}}$), then apply a correction for the deviation from equipartition to derive estimates of $R$ and $E$. For this deviation from equipartition, we assume that the total fraction of energy in the magnetic field is 2$\%$, based on observations of TDEs \citep{Cendes2021,Horesh2013} and supernovae \citep[e.g.][]{Eftekhari2018}. Additionally, we assume that a fraction of the total energy is carried by the electrons, as typically electrons are accelerated much less efficiently than protons in astrophysical accelerators \citep[e.g.][]{Morlino2012}. This fraction has frequently been assumed to be 10\%, i.e. $\epsilon_e=0.1$, in the literature \citep[e.g.][]{Alexander2016,Cendes2021,Goodwin2022} however, recent studies have found that for non-relativistic collisionless shocks, $\epsilon_e$ is closer to $10^{-3}-10^{-4}$ \citep{Xu2020,Park2015}. Thus, in our models we provide the outflow parameters for both $\epsilon_e=0.1$ and $\epsilon_e=5\times10^{-4}$. We note that the assumption of deviation from equipartition results in an increase in the predicted energy and radius as the assumed deviation from equipartition increases, as well as increased magnetic field strength and mass in the outflow with decreased ambient density. The modelled values depend on the assumed value of $\epsilon_e$, as demonstrated in Figure \ref{fig:outflowmodelling}. In Table \ref{tab:outflowproperties} we report both the equipartition radius and energy as well as the corrected radius and energy. 

We provide constraints for two different geometries in order to account for different possible outflow natures. First, we assume the emitting region is approximately spherical (with geometric factors\footnote{The geometric factors are given by $f_{\mathrm{A}}=A/(\pi R^2/\Gamma^2)$ and $f_{\mathrm{v}} = V/(\pi R^3/\Gamma^4)$, for area, $A$, and volume, $V$, of the outflow, and distance from the origin of the outflow, $R$ \citep{BarniolDuran2013}.} $f_{\mathrm{A}}=1$ and $f_{\mathrm{V}}=4/3$). Second, we provide outflow constraints assuming the emitting region is approximately conical (with geometric factors $f_{\mathrm{A}}=0.13$ and $f_{\mathrm{V}}=1.15$), corresponding to a mildly collimated jet with a half-opening angle of 30\,degrees. 

The inferred physical outflow properties for AT2020vwl are plotted in Figure \ref{fig:outflowmodelling} and listed in Table \ref{tab:outflowproperties}. A linear fit to the radius (assuming constant velocity) gives a predicted outflow launch date of -13$\pm$3\,d (spherical) or -13$\pm$2\,d (conical) after the initial optical detection. Due to the 18\,d uncertainty on the beginning of the optical flare, these predicted launch dates coincide with the predicted date of the onset of the optical flare.

The level of scintillation observed in the December epoch of observations (Section \ref{sec:scintillation}) provides an independent constraint on the source size to be likely $<134$\,microarcseconds (the size below which the object can be assumed to act as a point source affected by ISS) and definitely $<3028$\,microarcseconds (the maximum source size that could induce a 25\% modulation due to ISS), which is consistent with the radii predicted by the synchrotron modelling ($<93$ microarcseconds).

\section{Discussion}\label{sec:discussion}
The radio observations of the TDE AT2020vwl we present indicate that a non-relativistic outflow was launched approximately at the time of the initial optical flare. We deduce that the outflow has energy $\sim10^{48}$\,erg, for radii $\sim10^{16}$--$10^{17}$\,cm (assuming the correction for deviation from equipartition) and velocity $\approx0.03 \, c$ (spherical geometry) or $\approx0.07 \, c$ (conical geometry). The radio emission gradually faded over the course of our observations from 142--577\,d post-optical detection, with a slight indication of an increase in the peak flux density at 577\,d.  

\subsection{The nature of the outflow}
The exact mechanism behind radio emission in thermal TDEs is not known, including if all outflows are driven by the same mechanism, or if different mechanisms are behind different observed properties. Theoretical simulations predict slow ($\sim0.05-0.1\,c$), dispersed, spherical outflows from stream-stream collisions \citep[e.g.][]{Bonnerot2020} that would appear promptly after the initial optical detection and stellar disruption (our spherical model case in Figure~\ref{fig:outflowmodelling}). Alternative simulations predict non-relativistic, collimated radio outflows may be produced by the unbound debris stream \citep[e.g.][]{Spaulding2022}. Other theories suggest that the non-relativistic outflows could be driven by super-Eddington accretion induced winds from closer to the SMBH \citep[e.g.][]{Alexander2016} or a mildly collimated jet from accretion onto the SMBH \citep[e.g.][]{Pasham2018,Stein2021} (our conical model case in Figure~\ref{fig:outflowmodelling}). \citet{Dai2018} argue that wide-angle optically-thick fast outflows and relativistic jets are produced in the super-Eddington compact disk phase of TDEs, and the observed emission is highly dependent on viewing-angle. \citet{Curd2019} found that the combination of a rapidly spinning black hole and and strong magnetic field was required to launch a relativistic jet, but that a sub-relativisitic outflow could be driven by an accretion 'wind' due to low binding energy in the disk enabling small perturbations to unbind material relatively easily, driving an outflow due to radiation pressure. In many of these scenarios, the observable radio emission would appear quite similar, making it difficult to distinguish between different outflow scenarios. 

The outflow modelling we conducted in Section \ref{sec:modelling} enables some discrimination between outflow models. Firstly, the inferred radius of the outflow at each epoch enables us to rule out relativistic motion of the outflow between 142--577\,d post-optical detection. Secondly, a linear fit to the radial growth predicts an outflow launch date consistent with the optical detection within 1-$\sigma$ errors. Thirdly, the evolution of the predicted energy (corrected for any deviation from equipartition) in the outflow between epochs implies either fluctuating energy injection into the outflow from its source, or fluctuations in the density of the CNM the outflow is moving into.

Based on these observed properties, we deduce that the outflow from AT2020vwl is consistent with either a single injection of energy at the time of the stellar disruption and an inhomogenous CNM, or an energy source with fluctuating energy input (Figure \ref{fig:outflowmodelling}). The velocity over time is consistent with being approximately constant within the 1$\sigma$ error, whilst the mass in the ejecta gradually increased over time. Interestingly, the energy and radius increased with time until 307\,d (epoch 3) post-disruption, at which time the radius plateaued and the energy decreased for $\approx250$\,d, until a renewed increase in radius and energy was observed at 577\,d. This decrease in energy and stagnation of the radial growth are also reflected in the other parameters as a slight decrease in velocity and stagnation of mass in the outflow. We suggest that the decrease in energy of the outflow could be due to either the engine switching off (and switching back on at 577\,d), or, the outflow encountering a denser region of the CNM, slowing the blastwave. In the latter case, increasing energy in the outflow is due to the additional mass the outflow sweeps up from the CNM while undergoing ballistic motion from a single injection of energy at the time the outflow was launched. We note that in Figure \ref{fig:outflowmodelling} the ambient density appears to consistently decrease between each epoch, and shows no indication of a sudden increase in the final epoch which may explain the increase in energy. However, the error bars on the ambient density are large and cannot rule out the possibility that the ambient density increased in the final epoch. 

The prompt production of radio emission in this event (a linear fit to the radius gives an outflow launch date consistent with the time of the initial optical detection, also taking into account the 18\,d uncertainty on the onset of the optical flare), and lack of strong X-ray emission to indicate active accretion onto the SMBH, could indicate that an accretion-induced wind or jet producing the outflow from the vicinity of the SMBH is unlikely. However, if accretion were to occur promptly, the X-rays produced could be obscured or absorbed, and the radio outflow driven by material ejected from close to the SMBH. Additionally, infrared dust echos have been observed in TDEs that trace reverberation by gas that orbits the black hole \citep{vanvelzen2021b,Jiang2021}. The detection of an infrared dust echo could therefore be an indirect tracer of accretion occurring. We searched available Wide-field Infrared Survey Explorer \citep{Wright2010} observations of AT2020vwl post-flare and found there was no detectable increase in the infrared emission. Nevertheless, we cannot entirely rule out an accretion-driven wind as an explanation of AT2020vwl's radio emission.

The unbound debris stream could also produce the observed non-relativistic radio outflow. The unbound debris stream is predicted to have velocity $\approx 0.05 \, c$ and energy $\sim10^{48}$\,erg \citep{Krolik2016}, very similar to the properties we determine for AT2020vwl. However, we note that the unbound debris stream is likely to have a very small solid angle with higher collimation than we consider in our conical model ($f_A\approx0.2$). In this case, the predicted energies and velocities for our outflow modelling would increase slightly, and the outflow would be slightly more energetic and faster than predicted for the unbound debris stream. 

Stream-stream collisions during the initial disruption and circularisation of the debris could explain both the prompt ejection of the material in the outflow as well as the velocity and energy of the outflow that was observed for AT2020vwl. In this scenario, a collision induced outflow is produced by the self-intersection of the fallback stream, producing a prompt non-relativistic, spherical outflow with velocities $\lesssim0.2\,c$ \citep{Lu2020}. Similarly, a wide-angle outflow with velocity $\sim0.02\,c$ could be driven by shocks between returning streams and a circularising, eccentric accretion flow \citep{Steinberg2022}.

We therefore conclude that the outflow can either be explained by the unbound debris stream (less likely due to the higher collimation required) or more likely a spherical outflow that was launched by stream-stream collisions of the stellar debris. An accretion-driven outflow is also possible, but only if there was direct disk formation and obscuration of any X-ray emission. 

\subsection{Comparison to other TDEs}
\begin{figure*}
	\includegraphics[width=\columnwidth]{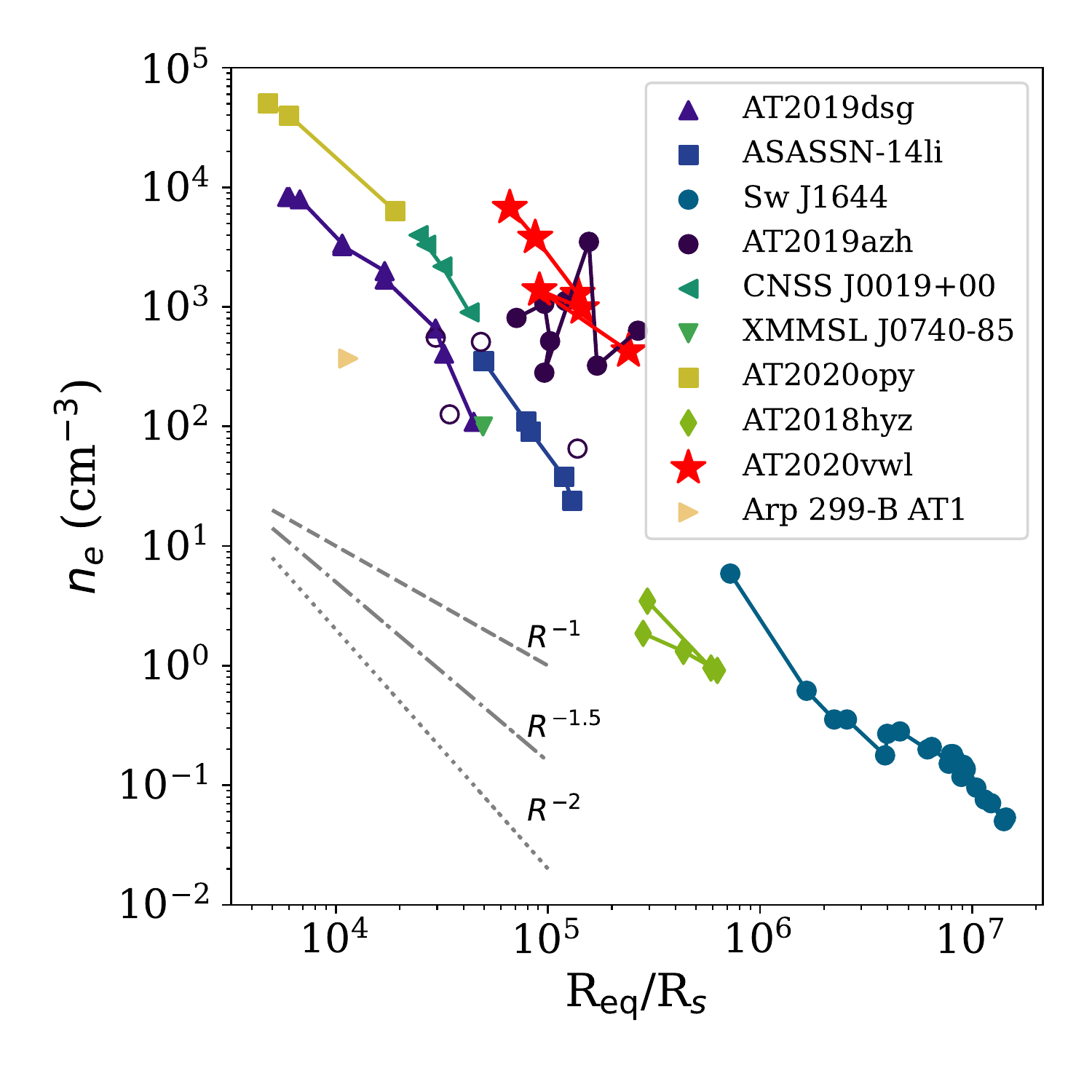}
	\includegraphics[width=\columnwidth]{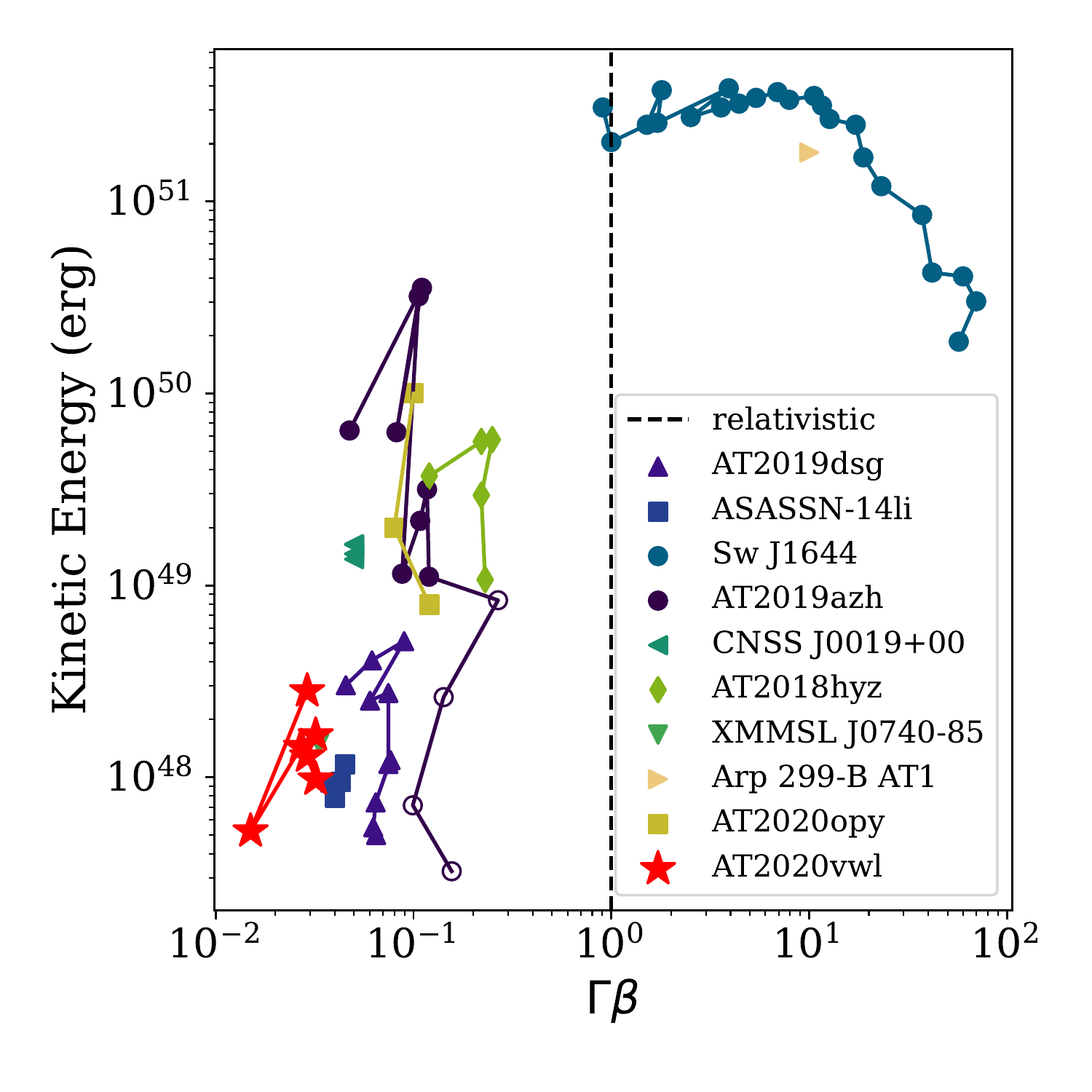}
    \caption{\textit{Left:} The variation of ambient density with distance from the black hole for a selection of thermal TDEs as traced by outflow modelling. The equipartition radius is plotted for AT2020vwl. \textit{Right:} The kinetic energy and velocity of the outflow produced in a selection of thermal TDEs. The equipartition corrected estimated kinetic energy is plotted for AT2020vwl. We plot the outflow properties inferred using $\epsilon_e=0.1$ for direct comparison to other inferred TDE outflow properties that are mostly calculated with $\epsilon_e=0.1$}. In both panels AT2020vwl is shown with red stars. AT2020vwl appears to fit well into the population of thermal TDEs.  TDE data and assumed SMBH masses are from \citet{Cendes2021,Stein2021} (AT2019dsg, $M_{\mathrm{BH}}=5\times10^6$\,$M_{\mathrm{\odot}}$), \citet{Alexander2016} (ASASSN-14li, $M_{\mathrm{BH}}=1\times10^6$\,$M_{\mathrm{\odot}}$), \citet{Eftekhari2018} (Sw J1644+57, $M_{\mathrm{BH}}=1\times10^6$\,$M_{\mathrm{\odot}}$), \citet{Anderson2020} (CNSS J0019+00, $M_{\mathrm{BH}}=1\times10^7$\,$M_{\mathrm{\odot}}$), \citet{Mattila2018} (Arp 299-B AT1, $M_{\mathrm{BH}}=2\times10^7$\,$M_{\mathrm{\odot}}$), \citet{Alexander2017} (XMMSL1 J0740-85, $M_{\mathrm{BH}}=3.5\times10^6$\,$M_{\mathrm{\odot}}$), \citet{Goodwin2022} (AT2019azh, $M_{\mathrm{BH}}=3\times10^6$\,$M_{\mathrm{\odot}}$), and \citet{Goodwin2023} (AT2020opy, $M_{\mathrm{BH}}=1.12\times10^7$\,$M_{\mathrm{\odot}}$). For AT2020vwl we assume  $M_{\mathrm{BH}}=6.17\times10^5$\,$M_{\mathrm{\odot}}$ \citep{Yao2023}.
    \label{fig:TDEcomparison}
\end{figure*}

In Figure \ref{fig:TDEcomparison} we plot the radius and ambient density of a number of TDEs as well as the kinetic energy and velocity of the outflows. The outflow properties of AT2020vwl are broadly consistent with those of other non-relativistic TDEs.  AT2020vwl is the slowest outflow observed to date (assuming a spherical geometry), with a velocity just $\approx 0.03 \, c$. 

The radio lightcurve evolution of AT2020vwl is very similar to the thermal TDE ASASSN-14li (Figure~\ref{fig:lightcurve}), in which the outflow was suggested to be produced by either a spherical, non-relativistic accretion-induced wind \citep{Alexander2016} or a more collimated, jet-like outflow \citep{vanVelzen2016,Pasham2018}. We observed AT2020vwl at radii closer to the central black hole than ASASSN-14li (Figure~\ref{fig:TDEcomparison}), possibly explaining the higher energy and denser CNM inferred from the synchrotron emission. The outflow launch date of AT2020vwl is consistent with the date of the initial optical detection. \citet{Alexander2016} found that the outflow launched by ASASSN-14li was launched approximately coincident with the onset of super-Eddington accretion and concluded that the outflow was likely launched by an accretion-driven wind. However, the initial optical flare and peak were not observed for ASAASN-14li, so the time of peak accretion may not coincide with the optical peak in this event, and the outflow could well have been launched prior to the onset of significant accretion onto the black hole, similar to AT2020vwl. Another well-studied non-relativistic TDE, AT2019dsg, was also found to have a radio outflow launched close to the time of optical discovery \citep{Cendes2021}.

The thermal TDES AT2019azh and AT2020opy demonstrated similar outflow properties (but slightly more energetic) to AT2020vwl, as well as predicted outflow launch dates that are also consistent with the optical flare \citep{Goodwin2022,Goodwin2023}, suggesting that these outflows may all have been produced by a similar mechanism. \citet{Goodwin2022} conclude that the most likely outflow mechanism for AT2019azh is ejected material due to stream-stream collisions of the stellar debris; a scenario which also explains the observed properties of AT2020vwl well. 

\citet{Steinberg2022} recently found in detailed simulations that the initial lightcurve rise is powered by shocks due to inefficient circularisation of the debris, and \citet{Metzger2022} found that significant accretion onto the SMBH is delayed as the envelope contracts slowly. These simulations lend credit to the collision-induced outflow scenario in which these prompt radio outflows such as observed for AT2020vwl, AT2019azh, AT2020opy, AT2019dsg, and perhaps ASASSN-14li are powered by shocks from the debris circularising, and not from accreted material close to the SMBH.

\section{Conclusions}\label{sec:conclusion}

We present the radio detection of the TDE AT2020vwl, and the evolution of the radio outflow that was produced over 1.5\,yr. The optical and radio lightcurves of this event are well-sampled, providing insight into the evolution timescales of the emission at different frequencies. We infer that the outflow in this event is non-relativistic, with velocity $\approx0.03 \, c$ (assuming a spherical outflow geometry), radius $10^{16}$--$10^{17}$\,cm and energy $10^{47}$--$10^{48}$\,erg. In VLA and uGMRT observations spaced 9\,d apart, we detected a 25$\%$ variation in flux density of the source which we attribute to interstellar scintillation, confirming the compact ($<134$\,microarcsecond) nature of the radio source. We deduce that the outflow in this event was likely produced by stream-stream collisions of the stellar debris, the unbound debris stream, or an accretion-induced outflow, due to the prompt onset of the radio emission relative to the optical flare and broadly constant low velocity of the outflow. We inferred an interesting decrease in the energy from 300-430\,d post-optical detection, possibly due to the outflow encountering a dense clump of the CNM or fluctuations in the energy injection into the outflow. 

Future radio observations that track the decay of the radio emission from this source will enable stronger constraints on the host component of the radio emission, as well as the evolution of the outflow and density of the CNM further from the central SMBH. AT2020vwl joins a growing number of TDEs with prompt radio emission detected relative to the optical flare, motivating future prompt radio observations combined with detailed optical and X-ray observations. These observations may enable the mechanism behind these outflows to be distinguished from accretion-driven ejection from close to the SMBH or collision driven outflowing streams of material. More detailed simulations of the launching and evolution timescales of these different types of outflows may also provide key insights into unveiling the nature of prompt outflows from TDEs.

\section*{Acknowledgements}
We thank the anonymous referee for insightful comments that helped to improve the manuscript. This work was supported by the Australian government through the Australian Research Council’s Discovery Projects funding scheme (DP200102471). The National Radio Astronomy Observatory is a facility of the National Science Foundation operated under cooperative agreement by Associated Universities, Inc. We thank the staff of the GMRT that made these observations possible. GMRT is run by the National Centre for Radio Astrophysics of the Tata Institute of Fundamental Research. The MeerKAT telescope is operated by the South African Radio Astronomy Observatory, which is a facility of the National Research Foundation, an agency of the Department of Science
and Innovation. T.E. is supported by NASA through the NASA Hubble Fellowship grant HST-HF2-51504.001-A awarded by the Space Telescope Science Institute, which is operated by the Association of Universities for Research in Astronomy, Inc., for NASA, under contract NAS5-26555.

\section*{Data Availability}

The radio observations presented in Table \ref{tab:radio_obs} will be available online in machine readable format.



\bibliographystyle{mnras}
\bibliography{bibfile,mfb_temp} 



\appendix

\section{Spectral fitting}\label{sec:Appendix}

The observed radio spectra of AT2020vwl show a peaked synchrotron spectrum that evolved to peaking at lower frequencies over time. In Section \ref{sec:specfits} we present spectral fits assuming the synchrotron spectrum is described by the case where $\nu_m<\nu_a<\nu_c$, and we subtract a constant steep-spectrum faint host component from each epoch. Here we explore alternative scenarios that could produce the synchrotron emission that was observed, noting that the spectral fits presented in the main body of the paper provide the statistically best fit to the data. 

In order to assess the statistically best spectral fit, here we compare multiple models using the Akaike's information criterion (AIC) and the Schwarz Bayesian information criterion (BIC). The AIC and BIC are calculated as follows
\begin{equation}
    AIC = - 2L + 2q
\end{equation}

\begin{equation}
    BIC = - 2L + q ln(N)
\end{equation}
where $L$ is the log-likelihood, $q$ is the number of fit parameters and $N$ is the total number of data points. 

A lower AIC and BIC indicates a "better" fit, with a statistically better fit defined by an AIC or BIC 2 points lower than the alternate model. 

\subsection{The shape of the observed synchrotron spectrum}

\subsubsection{Fits with a single break}

Here we assess whether the observed synchrotron spectra are better described by a single break, where the break frequency is associated with either the synchrotron self-absorption frequency, or the minimum frequency. 

The optically thick component of the observed synchrotron spectra is described by $\nu^{5/2}$ in the case in which the peak of the spectrum is associated with the synchrotron self-absorption break, whereas it is described by $\nu^{1/3}$ in the case in which the peak is associated with the minimum frequency break (i.e. the case in which $\nu_a<\nu_m<\nu_c$).

In order to assess the regime that best describes the observed spectra of AT2020vwl, we fit a power law to the optically thick slope of the first epoch (before subtracting any host emission), 2020 February 27, in which the optically thick portion of the spectrum is the best-constrained. This fit produced $\nu^{1.6\pm0.3}$, i.e.\ between the $\nu^{1/3}$ and $\nu^{5/2}$ expected for the minimum frequency or self-absorption breaks respectively. To further assess the possibility that the synchrotron emission is in the regime $\nu_a<\nu_m<\nu_c$, we fit all epochs both with and without the host component assuming the synchrotron peak is associated with the minimum frequency, i.e.

\begin{equation}
    \label{eq:Fnum}
    \begin{aligned}
        F_{\nu, \mathrm{synch}} = F_{\nu,\mathrm{ext}} \left[
        \left(\frac{\nu}{\nu_{\rm m}}\right)^{-s\beta_1} +  \left(
        \frac{\nu}{\nu_{\rm m}}\right)^{-s\beta_2)
        }\right]^{-1/s}
        \end{aligned}
    \end{equation}
    where $\nu$ is the frequency, $F_{\nu,\mathrm{ext}}$ is the normalisation, $s = 1.84-0.4p$, $\beta_1 = \frac{1}{3}$, $\beta_2 = \frac{1-p}{2}$. 
    
As well as the case where the synchrotron peak is associated with the self-absorption frequency, i.e.

\begin{equation}
    \label{eq:Fnua}
    \begin{aligned}
        F_{\nu, \mathrm{synch}} = F_{\nu,\mathrm{ext}} \left[
        \left(\frac{\nu}{\nu_{\rm a}}\right)^{-s\beta_1} +  \left(
        \frac{\nu}{\nu_{\rm a}}\right)^{-s\beta_2)
        }\right]^{-1/s}
        \end{aligned}
    \end{equation}
    where $\nu$ is the frequency, $F_{\nu,\mathrm{ext}}$ is the normalisation, $s = 1.25-0.18p$, $\beta_1 = \frac{5}{2}$, $\beta_2 = \frac{1-p}{2}$. 

The spectral fit for a single break where the break is associated with the minimum frequency resulted in a statistically worse fit to the observed data when compared to the break being associated with the self-absorption frequency for 4 out of 6 of the epochs (without host subtraction) and 5 out of 6 of the epochs (with host subtraction), with the two models compared in Table \ref{tab:specfits_AIK}.

\begin{table}
    \centering
\caption{AIC and BIC values for single-break spectral fits assuming the synchrotron peak frequency is associated with the minimum frequency ($\nu_m$) or the self-absorption frequency ($\nu_a$).}
    \begin{tabular}{c|cc|cc}
    \hline
Date (UTC) & AIC $\nu_m$ & AIC $\nu_a$ & BIC $\nu_m$ & BIC $\nu_a$\\
\hline
Without host subtraction \\
\hline

2021-02-27 &  -11.98 & -9.51 & -10.60 & -8.13 \\
2021-05-07 &  -14.35 & -13.58 & -13.79 & -13.03 \\
2021-08-11 &  -38.89 & -44.81 & -36.10 & -42.03 \\
2021-10-18 &  -42.81 & -47.97 & -39.41 & -44.58 \\
2021-12-14 &  -48.50 & -50.47 & -45.11 & -47.07 \\
2022-05-08 &  -43.95 & -47.73 & -39.48 & -43.25 \\

\hline
With host subtraction \\
\hline

2021-02-27 &  -9.09 & -14.51 & -7.71 & -13.13 \\
2021-05-07 &  -14.74 & -14.88 & -14.19 & -14.33 \\
2021-08-11 &  -41.65 & -47.03 & -38.87 & -44.24 \\
2021-10-18 &  -45.49 & -48.86 & -42.09 & -45.46 \\
2021-12-14 &  -49.89 & -51.59 & -46.50 & -48.19 \\
2022-05-08 &  -47.67 & -50.77 & -43.20 & -46.29 \\

\hline
    \end{tabular}
    \label{tab:specfits_AIK}
\end{table}

\subsubsection{Fits with two spectral breaks}

We note that a slightly flatter optically thick spectral slope can be obtained when $\nu_m$ and $\nu_a$ are close, which also gives a broader peak. In our best-fit models, $\nu_m$ and $\nu_a$ are both fit for, and in each epoch the break frequencies are always within 4\,GHz of each other (Table \ref{tab:outflowproperties}). Here we assess whether a single break model (Equation \ref{eq:Fnua}), in which only the $\nu_a$ break is present in the spectrum, or a two-break model (Equation \ref{eq:Fv}), in which both $\nu_a$ and $\nu_m$ are present in the spectrum, provides a statistically better fit.

\begin{table}
    \centering
\caption{AIC and BIC values for two-break spectral fits assuming the synchrotron peak frequency is associated with the self-absorption frequency ($\nu_a$) and including a second minimum frequency break ($\nu_m$) that is fixed to $\nu_m=0.5$\,GHz or also allowed to vary.}
    \begin{tabular}{p{2.5cm}|p{1cm}p{1cm}|p{1cm}p{1cm}}
    \hline
Date (UTC) & AIC fixed $\nu_m$ break & AIC $\nu_a$ and $\nu_m$ breaks & BIC fixed $\nu_m$ break & BIC  $\nu_a$ and $\nu_m$ breaks\\
\hline
Without host subtraction \\
\hline

2021-02-27 &  -21.93 & -21.98 & -20.55 & -20.60 \\
2021-05-07 &  -29.69 & -26.69 & -29.14 & -29.14 \\
2021-08-11 &  -48.34 & -48.34 & -45.56 & -45.56 \\
2021-10-18 &  -48.86 & -48.87 & -45.47 & -45.47 \\
2021-12-14 &  -51.54 & -51.54 & -48.14 & -48.14 \\
2022-05-08 &  -50.81 & -50.81 & 46.34- & -46.34 \\

\hline
With host subtraction \\
\hline

2021-02-27 &  -25.84 & -25.86 & -24.46 & -24.48 \\
2021-05-07 &  -29.80 & -29.80 & -29.25 & -29.25 \\
2021-08-11 &  -48.17 & -48.17 & -45.38 & -45.38 \\
2021-10-18 &  -49.03 & -49.03 & -45.63 & -45.64 \\
2021-12-14 &  -51.68 & -51.68 & -48.28 & -48.28 \\
2022-05-08 &  -50.90 & -50.91 & -46.43 & -46.43 \\

\hline
    \end{tabular}
    \label{tab:specfits_breaks_AIK}
\end{table}

The AIC and BIC for the two break models (fixed $\nu_m$ or fit $\nu_m$) are reported in Table \ref{tab:specfits_breaks_AIK}. It is clear that a two break model with $\nu_m$ fixed or allowed to vary are equally statistically preferred for all epochs. Comparing the AIC and BIC values with and without host subtraction for the one or two break models (Tables \ref{tab:specfits_AIK} and \ref{tab:specfits_breaks_AIK}), it is clear that a two break model is statistically preferred in all 6 epochs. 

\subsection{Spectral fits with and without a host component}

Additionally, in order to assess whether the spectral fits including a host component are statistically better than those without a host component, here we compare the two models for the preferred case in which $\nu_a<\nu_m<\nu_c$. 

In Table \ref{tab:specfits_breaks_AIK} it is clear that the AIC and BIC indicate the model including a host component gives a statistically better fit to the data for the first epoch, and show no preference for epochs 2--6 with very similar AIC and BIC values for both models. We thus conclude that in general the data is equally well-fit by the model with or without a host component. We note that the first epoch is particularly susceptible to host contamination as the synchrotron spectrum peak flux density is above 6\,GHz and the host emission is greater at lower frequencies. For this reason, we decided to include a host component (assumed to be constant) in our spectral fits.  

\begin{figure}
	\includegraphics[width=\columnwidth]{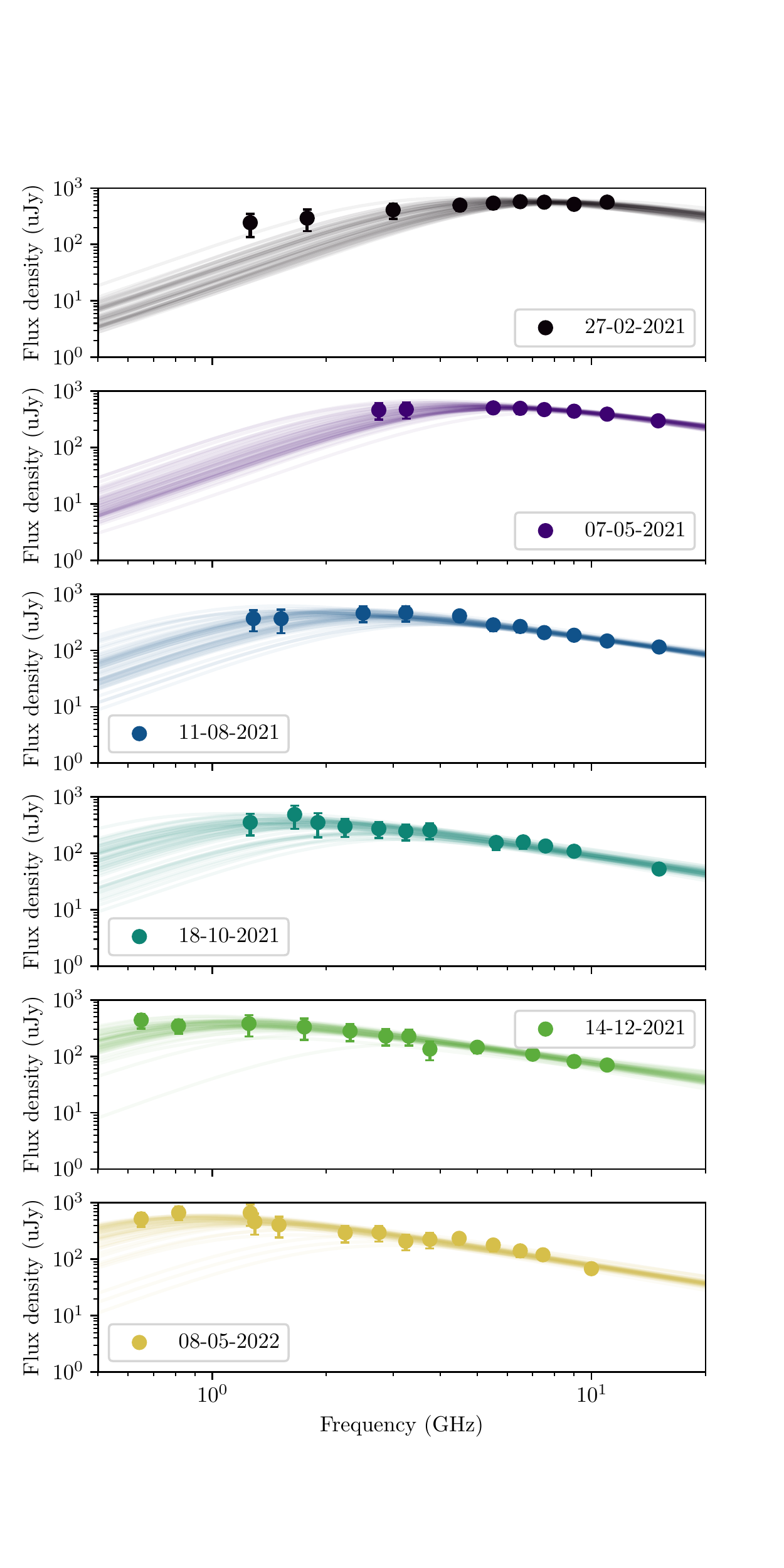}
    \caption{Synchrotron spectral fits of the evolving radio emission observed from AT2020vwl between 2020 February to 2022 May. The observed radio flux densities without subtracting any host component are plotted in circles and 50 random MCMC samples are plotted in solid lines to demonstrate the uncertainty in the fits. The radio emission is well described by a synchrotron self-absorption spectrum that evolves to peaking at lower frequencies over time.}
    \label{fig:nohost_specfits}
\end{figure}

For completeness, here we also include the fitted spectral and corresponding outflow properties under the assumption of no contaminating host emission in the total radio flux density measured. In Figure \ref{fig:nohost_specfits} and Table \ref{tab:nohost}, we give the spectral fits and equipartition model parameters assuming no host component in the spectral fits.

\begin{table*}
	\centering
	\caption{Synchrotron emission and equipartition modelling properties of the outflow produced by the TDE AT2020vwl assuming no host emission contribution in the spectral fitting.}
	\label{tab:nohost}
 \begin{tabular}{p{1.5cm}p{1cm}p{1cm}p{1.1cm}p{1.1cm}p{1.3cm}p{1.1cm}p{1.1cm}p{1.1cm}p{1.1cm}p{1.3cm}p{1.3cm}}
        \hline
Date (UTC) & $\delta$t (d)$^{a}$ & $F_{\rm{p}}$ ($\mu$Jy) & $\nu_{\rm{m}}$ (GHz)& $\nu_{\rm{p}}$ (GHz) & $p$ & log$E$ (erg) &  log$R$ (cm) &
 $\beta$ (c) & log$n_{e}$ (cm$^{-3}$) & log$B$ (G) & log$M_{\rm ej}$ (M$_{\odot}$) \\
	\hline
Spherical, $\epsilon_e=0.1$ & & & & & & & & & & & \\
\hline
27-02-2021 & 142 &
0.66$\pm$0.05& 3.17$\pm$0.93 & 7.04$\pm$0.81 & 2.82$\pm$    0.27 &
48.07$\pm$0.07 & 16.15$\pm$0.06 &    0.039$\pm$0.005 &     3.68$\pm$0.49 &     -0.20$\pm$0.38 &     -3.05$\pm$0.09 \\
07-05-2021 & 211 &
0.61$\pm$0.05& 2.35$\pm$0.78 & 5.44$\pm$0.65 & 2.86$\pm$    0.26 &
48.18$\pm$0.08 & 16.25$\pm$0.06 &    0.033$\pm$0.005 &     3.50$\pm$0.54 &     -0.29$\pm$0.41 &     -2.80$\pm$0.10 \\
11-08-2021 & 307 &
0.50$\pm$0.08& 1.13$\pm$0.39 & 2.45$\pm$0.38 & 2.96$\pm$    0.22 &
48.50$\pm$0.11 & 16.57$\pm$0.08 &    0.046$\pm$0.008 &     2.88$\pm$0.67 &     -0.60$\pm$0.49 &     -2.77$\pm$0.14 \\
18-10-2021 & 375 &
0.38$\pm$0.08& 0.96$\pm$0.34 & 1.85$\pm$0.34 & 3.10$\pm$    0.19 &
48.59$\pm$0.14 & 16.64$\pm$0.09 &    0.044$\pm$0.009 &     2.75$\pm$0.83 &     -0.67$\pm$0.58 &     -2.66$\pm$0.17 \\
14-12-2021 & 432 &
0.41$\pm$0.06& 0.63$\pm$0.13 & 1.20$\pm$0.17 & 2.93$\pm$    0.24 &
48.68$\pm$0.10 & 16.83$\pm$0.06 &    0.059$\pm$0.008 &     2.26$\pm$0.50 &     -0.91$\pm$0.36 &     -2.82$\pm$0.11 \\
08-05-2022 & 577 &
0.55$\pm$0.10& 0.60$\pm$0.10 & 1.05$\pm$0.13 & 3.13$\pm$    0.16 &
49.05$\pm$0.11 & 16.96$\pm$0.06 &    0.060$\pm$0.008 &     2.25$\pm$0.54 &     -0.92$\pm$0.36 &     -2.45$\pm$0.12 \\
\hline
Conical, $\epsilon_e=0.1$ & & & & & & & & & & & \\
\hline

& & &  &  &
 & 48.59$\pm$0.07 & 16.53$\pm$0.06 &    0.087$\pm$0.012 &     3.14$\pm$0.49 &     -0.47$\pm$0.38 &     -3.24$\pm$0.09 \\
& & &  &  &
& 48.70$\pm$0.08 & 16.63$\pm$0.06 &    0.074$\pm$0.011 &     2.96$\pm$0.54 &     -0.56$\pm$0.41 &     -2.99$\pm$0.10 \\
& & &  &  &
 & 49.03$\pm$0.11 & 16.94$\pm$0.08 &    0.103$\pm$0.018 &     2.35$\pm$0.67 &     -0.87$\pm$0.49 &     -2.94$\pm$0.14 \\
& & &  &  &
 & 49.13$\pm$0.14 & 17.02$\pm$0.09 &    0.100$\pm$0.021 &     2.23$\pm$0.83 &     -0.93$\pm$0.58 &     -2.82$\pm$0.17 \\
& & &  &  &
 & 49.21$\pm$0.10 & 17.21$\pm$0.06 &    0.130$\pm$0.017 &     1.72$\pm$0.50 &     -1.18$\pm$0.36 &     -2.97$\pm$0.11 \\
& & &  &  &
& 49.59$\pm$0.11 & 17.34$\pm$0.06 &    0.132$\pm$0.018 &     1.72$\pm$0.54 &     -1.18$\pm$0.36 &     -2.60$\pm$0.12 \\

\hline
\hline
	\end{tabular}
	\footnotesize{$^{a}$ $\delta$t is measured with respect to the initial optical detection, $t_0=$ MJD 59130}
\end{table*}

When no host component is accounted for, the resulting fitting and equipartition analysis infer a slightly larger, faster, and more energetic outflow than when a small amount of host emission is assumed. The overall trend in the outflow properties remains the same. 


\bsp	
\label{lastpage}
\end{document}